\documentclass[12pt,aps,manuscript]{revtex4-1}
\usepackage{lmodern}
\usepackage{lmodern}
\usepackage[T1]{fontenc}
\usepackage[utf8]{inputenc}
\usepackage[a4paper]{geometry}
\usepackage{amsmath}
\usepackage{amssymb}
\usepackage{graphicx}

\makeatletter

\providecommand{\tabularnewline}{\\}

\@ifundefined{textcolor}{}
{%
 \definecolor{BLACK}{gray}{0}
 \definecolor{WHITE}{gray}{1}
 \definecolor{RED}{rgb}{1,0,0}
 \definecolor{GREEN}{rgb}{0,1,0}
 \definecolor{BLUE}{rgb}{0,0,1}
 \definecolor{CYAN}{cmyk}{1,0,0,0}
 \definecolor{MAGENTA}{cmyk}{0,1,0,0}
 \definecolor{YELLOW}{cmyk}{0,0,1,0}
}
\numberwithin{equation}{section}

\usepackage{amsmath}
\DeclareMathOperator{\Tr}{Tr}
\DeclareMathOperator{\sgn}{sgn}
\DeclareMathOperator{\TDL}{TDL}

\usepackage[titletoc,title]{appendix}

\usepackage{braket}

\makeatother

\begin{document}

\title{Quenching the XXZ spin chain: quench action approach versus generalized
Gibbs ensemble}

\author{M. Mestyán$^{1}$, B. Pozsgay$^{1}$, G. Takács$^{1}$ and M. A.
Werner$^{2}$ }

\affiliation{$^{1}$MTA-BME “Momentum” Statistical Field Theory Research Group
\\
Department of Theoretical Physics, Budapest University of Technology
and Economics\\
1111 Budapest, Budafoki út 8, Hungary \\
$^{2}$MTA-BME “Momentum” Exotic Quantum Phases Research Group \\
Department of Theoretical Physics, Budapest University of Technology
and Economics \\
1111 Budapest, Budafoki út 8, Hungary}

\date{15th December 2014}
\begin{abstract}
Following our previous work {[}PRL 113 (2014) 09020{]} we present
here a detailed comparison of the quench action approach and the predictions
of the generalized Gibbs ensemble, with the result that while the
quench action formalism correctly captures the steady state, the GGE
does not give a correct description of local short-distance correlation
functions. We extend our studies to include another initial state,
the so-called q-dimer state. We present important details of our construction,
including new results concerning exact overlaps for the dimer and
q-dimer states, and we also give an exact solution of the quench-action-based
overlap-TBA for the q-dimer. Furthermore, we extend our computations
to include the $xx$ spin correlations besides the $zz$ correlations
treated previously, and give a detailed discussion of the underlying
reasons for the failure of the GGE, especially in the light of new
developments.
\end{abstract}
\maketitle

\section{Introduction}

Non-equilibrium evolution of isolated quantum many-body systems has
recently come to the center of attention \cite{2011RvMP...83..863P,2007PhRvL..98r0601K,2008Natur.452..854R,2008PhRvL.100j0601B,2009PhRvL.103j0403R,2010NJPh...12e5020C,2010PhRvE..81c6206S,2010NJPh...12e5017B,2011PhRvL.106e0405B,2011PhRvL.106v7203C,2012JSMTE..07..016C,2013PhRvL.110j0406S,2013PhRvL.111s7203M,2013PhRvA..87e3628I}
due to spectacular recent advances experiments with ultra-cold atoms
\cite{2006Natur.440..900K,2012Natur.481..484C,2012NatPh...8..325T,2013Natur.502...76F}.
Whether isolated quantum systems reach an equilibrium in some appropriate
sense, and, if the answer is yes, the nature of the steady state reached,
are long-standing and fundamental problems in theoretical physics.

For generic systems it is expected that provided driving forces are
absent, after a sufficiently long time they reach a steady state in
which the expectation values of some class of relevant observables
are described by a thermal Gibbs ensemble \cite{2008Natur.452..854R,2011RvMP...83..863P}.
The choice of the class of observables generally follows the idea
that they are supported on subsystems which in the thermodynamic limit
are infinitely smaller than the rest of the system. The rest of the
system can then act as a heat bath, leading to thermalization. Such
classes are given by local observables (e.g. short range correlators)
on a chain with local Hamiltonian, or observables involving (sums
over) few-body operators in a many-body system.

Thermalization, however, is only expected to hold for systems with
generic, i.e. non-integrable dynamics. Dynamics of integrable systems
is constrained by the conservation of extra charges which prevents
relaxation to a thermal state. It was suggested in \cite{2007PhRvL..98e0405R}
that in the integrable case the long-time asymptotic stationary state
is described by a statistical ensemble involving all the relevant
conserved charges $\{\hat{Q}_{i}\}$, the Generalized Gibbs Ensemble
(GGE). When considering local quantities as relevant observables,
it is intuitively clear that the relevant charges to include are the
local ones. In the case of integrable systems, the generating function
of such charges is a commuting family of transfer matrices as a function
of the so-called spectral parameter \cite{9780511628832}.

The GGE can be derived by applying the maximum entropy principle under
the constraint provided by the charges $\{\hat{Q}_{i}\}$, therefore
the idea is very natural in the framework of statistical mechanics.
However, it is quite difficult to construct the ensemble for strongly
correlated genuinely interacting quantum systems. Therefore most initial
studies of GGE were carried out in theories equivalent to free fermions
\cite{2006PhRvL..97o6403C,2009PhRvL.102l7204R,2012JSMTE..07..022C,2012PhRvE..85a1133C,2013JSMTE..02..014G,2013PhRvB..87x5107F,2014JSMTE..03..016F,2013PhRvL.110x5301C,2014PhRvA..89a3609K,2014JSMTE..07..024S}
or by numerical studies of relatively small systems \cite{2011PhRvL.106n0405C,2014PhRvL.113e0601W}.
More recently it became possible to examine genuinely interacting
integrable systems such as the 1D Bose gas \cite{2012PhRvL.109q5301C,2013PhRvB..88t5131K,2014PhRvA..89c3601D},
the XXZ Heisenberg spin chain \cite{2013JSMTE..07..003P,2013JSMTE..07..012F,2014PhRvB..89l5101F}
or field theories \cite{2010NJPh...12e5015F,2013PhRvL.111j0401M,2014PhLB..734...52S}.
Even so, it took quite some time until the first precision numerical
test of predictions of the GGE against real time dynamics was performed
\cite{2014PhRvB..89l5101F}. 

Surprisingly, the validity of the GGE for genuinely interacting theories
has been called into question by a series of recent studies. A crucial
step in this direction was the development of the quench action approach
\cite{Caux2013}, which provided an alternative way to study the time
evolution using the overlaps of the initial state with the eigenstates
of the post-quench Hamiltonian. In particular it allows to derive
overlap thermodynamic Bethe Ansatz (oTBA) equations for the steady
states, provided the exact overlaps are known. Using the results of
\cite{2012JSMTE..05..021K}, a determinant formula for overlaps with
the Néel state was first computed in \cite{Pozsgay2014a}. It was
substantially improved in \cite{2014JPhA...47n5003B,Brockmann2014c},
allowing the evaluation of the thermodynamic limit. In addition, the
dimer state overlaps were also expressed in terms of the Néel ones
in \cite{Pozsgay2014a}. The oTBA equations for the Néel state were
first obtained in \cite{Wouters2014}, and it was also shown that
the GGE and the oTBA give different results for the Bethe root densities,
and also for the nearest neighbour spin-spin correlators for the case
of the Néel initial state; the difference, however, is very small. 

The oTBA equations were also derived independently for the Néel and
dimer states in \cite{Pozsgay2014}, where we compared the GGE and
oTBA results to numerical real time evolution computed using the infinite-volume
Time-Evolving Block Decimation (iTEBD) method \cite{2004PhRvL..93d0502V,2007PhRvL..98g0201V}.
It turned out that while the precision of the iTEBD is not enough
to resolve the difference between the GGE and the oTBA for the Néel
state, in the dimer case the issue can be unambiguously decided: the
GGE built upon the local conserved charges fails to describe the local
correlators in the steady state, while the oTBA results agree perfectly
with the numerics. Three elements proved to be necessary to arrive
at a definite conclusion:
\begin{enumerate}
\item A nontrival and novel conjecture, published and thoroughly tested
in \cite{Mestyan2014}, which enabled the construction of local correlators
from the TBA solution (independently of whether it was derived for
a thermal or GGE ensemble, or from the quench action approach).
\item The exact overlaps of the dimer state, computed using the results
in \cite{Pozsgay2014a}.
\item Explicit numerical evaluation of the time evolution using the infinite-size
Time Evolving Block Decimation (iTEBD) algorithm developed in \cite{2004PhRvL..93d0502V,2007PhRvL..98g0201V}. 
\end{enumerate}
In a subsequent version of \cite{Wouters2014}, using the results
for correlators derived in \cite{Mestyan2014} it was also shown that
the oTBA reproduces the diagonal ensemble, while the GGE differs from
it.

In the present paper we present previously unpublished background
material behind the work \cite{Pozsgay2014}, such as the derivation
of the dimer overlaps, the details of the numerical time evolution,
and the results for the $xx$ spin correlators. We also extend our
results to a q-deformed version of the dimer state: we give the exact
overlaps, construct the oTBA and the GGE predictions, and compare
them to the iTEBD results. It turns out that the oTBA and GGE gives
different predictions, but again the difference is to small to be
resolved by the numerics; however, it provides an important consistency
check for our framework. We also show that the exact solution proposed
for the TBA equations in \cite{Wouters2014} can be extended to the
q-dimer case as well, and give some more intuitive background for
it by relating it to the Loschmidt echo studied in \cite{Pozsgay2013}.
In addition, we derive a partially decoupled version of the formulas
for the correlation functions. 

A further motivation for this paper is that the original results have
been widely discussed since its publication in \cite{Pozsgay2014},
and that some very important follow-up appeared which clarified some
interesting aspects of the failure of the GGE \cite{Goldstein2014,2014JSMTE..09..026P,Pozsgay2014c}.
We give an exposition and discussion of the issues and arguments in
our conclusions.

The outline of the paper is as follows. Section \ref{sec:Overview-of-the-XXZ-TBA}
gives a brief overview of the XXZ Bethe Ansatz, where we collect the
necessary facts and set up our notations. Section \ref{sec:Bethe-Ansatz-for-QQ-in-XXZ}
describes the application of the Bethe Ansatz to quantum quenches
in the XXZ chain. We briefly describe the GGE and how to compute thermodynamics
in the GGE using TBA formalism, and give a summary of the quench action
approach. In Section \ref{sec:steadyStateDensities} we go through
the exact overlaps and give their construction for the dimer and q-dimer
states. Then we turn to the oTBA, discuss its exact solution for the
Néel and extend it to the q-dimer case. Section \ref{sec:Computing-correlation-functions}
summarizes the methods necessary to compute the correlation functions
from the oTBA solutions, which are then compared to the iTEBD in Section
\ref{sec:Discussion}. 

Some longer derivations and technical details are relegated to appendices.
Appendix \ref{sec:appendixCorrelators} contains the derivation of
the decoupled correlator formulas, while Appendix \ref{sec:appendixTEBD}
describes the details of the iTEBD, including issues of reliability
and error estimation.

\section{Overview of the Bethe Ansatz for the \emph{XXZ }spin chain\label{sec:Overview-of-the-XXZ-TBA}}

The Hamiltonian of the XXZ spin chain is

\begin{equation}
H_{\textit{XXZ}}(\Delta)=\sum_{j=1}^{L}(\sigma_{j}^{x}\sigma_{j+1}^{x}+\sigma_{j}^{y}\sigma_{j+1}^{y}+\Delta(\sigma_{j}^{z}\sigma_{j+1}^{z}-1))\,,\label{eq:XXZHamiltonian}
\end{equation}
where $\Delta$ is the anisotropy parameter, and we impose periodic
boundary conditions $\sigma_{j+L}\equiv\sigma_{j}$. The Hamiltonian
can be diagonalized using the Bethe Ansatz \cite{Bethe1931,Orbach1958},
which we briefly summarize in the following in order to set down our
notations. Since the Hamiltonian conserves the total value of the
$z$ component of the spin 
\begin{equation}
[H,S^{z}]=0\qquad S^{z}=\sum_{j=1}^{L}\sigma_{j}^{z}\;,
\end{equation}
the eigenstates are of the form
\begin{equation}
\begin{aligned}|\{\lambda_{j}\}_{j=1}^{M}\rangle & =\sum_{n_{1}<n_{2}<...<n_{M}}\Psi(\{\lambda_{j}\}_{j=1}^{M}|n_{1},n_{2},...,n_{M})|n_{1},n_{2},...,n_{M}\rangle\\
|n_{1},n_{2},...,n_{M}\rangle & =\sigma_{n_{1}}^{-}\sigma_{n_{2}}^{-}...\sigma_{n_{M}}^{-}|\uparrow\uparrow...\uparrow\rangle\,,
\end{aligned}
\label{eq:basisExpansion}
\end{equation}
parametrized by rapidities $\{\lambda_{j}\}_{j=1}^{M}$, and having
$S^{z}=L/2-M$. The corresponding wave function is built from $M$
plane waves with factorized scattering amplitudes:
\begin{equation}
\Psi(\{\lambda_{j}\}_{j=1}^{M}|n_{1},n_{2},...,n_{M})=\sum_{\pi\in S_{M}}I(\pi)\prod_{j=1}^{M}\left(\frac{\sin(\lambda_{j}+i\eta/2)}{\sin(\lambda_{j}-i\eta/2)}\right)^{n_{\pi(j)}}\left(\prod_{j<k}\frac{\sin(\lambda_{\pi(j)}-\lambda_{\pi(k)}+i\eta)}{\sin(\lambda_{\pi(j)}-\lambda_{\pi(k)}-i\eta)}\right)\,,\label{eq:BetheWaveFunction}
\end{equation}
in which $\pi$ is a permutation of $1,\dots,M$ with parity $I(\pi)$,
and $\eta$ is defined by $\cosh\eta=\Delta$. The Bethe equations
follow from imposing the periodic boundary condition:
\begin{equation}
\left(\frac{\sin(\lambda_{j}+i\eta/2)}{\sin(\lambda_{j}-i\eta/2)}\right)^{L}=\prod_{k\neq j}\left(\frac{\lambda_{j}-\lambda_{k}+i\eta}{\lambda_{j}-\lambda_{k}-i\eta}\right)\,.\label{eq:BetheEquations}
\end{equation}
The thermodynamics of the chain can be described on the basis of the
string hypothesis; the exposition below follows the work \cite{9780511524332}.
The thermodynamic limit is defined by 
\begin{equation}
L,\, M\rightarrow\infty\qquad\mbox{ such that}\qquad M/L=\mbox{const.}\,,
\end{equation}
and shall be abbreviated by $\TDL$. For large but finite values of
the chain length $L$, the rapidities $\{\lambda_{j}\}_{j=1}^{M}$
parametrizing the wave function are organized into configurations
of approximate strings. The $j$-th rapidity in the $\alpha$-th approximate
$n$-string is given by 
\begin{equation}
\lambda_{\alpha,j}^{n}=\lambda_{\alpha}^{n}+\frac{i\eta}{2}(n+1-2j)+\delta_{\alpha,j}^{n}\,.
\end{equation}
The string hypothesis states that the deviations $\delta_{\alpha,j}^{n}$
are $\mathcal{O}(1/L)$, therefore the thermodynamic limit of the
wave function contains $n$-strings that are fully defined by their
real part $\lambda_{\alpha}^{n}$. Due to the above definition of
the thermodynamic limit, the strings will have a finite density in
rapidity space: the number of $n$-strings in a rapidity interval
$(\lambda,\lambda+d\lambda)$ is $L\rho_{n}(\lambda)d\lambda$, where
$\rho_{n}(\lambda)$ is the density of $n$-strings.

For any given set of string densities $\{\rho_{n}(\lambda)\}$, there
are usually many Bethe Ansatz eigenstates which scale to $\{\rho_{n}(\lambda)\}$.
However, we assume that the expectation values of relevant observables
are entirely determined by $\{\rho_{n}(\lambda)\}$ in the thermodynamic
limit; one can consider this as a selection condition for observables
to which the thermodynamic limit applies.

It is useful to define the entropy per site 
\begin{equation}
s[\{\rho_{n}(\lambda)\}]=\frac{1}{L}\ln\,\mathcal{N}[\{\rho_{n}(\lambda)\}]\,,\label{eq:entropyDefinition}
\end{equation}
where $\mathcal{N}[\{\rho_{n}(\lambda)\}]$ is the number of Bethe
Ansatz eigenstates scaling to $\{\rho_{n}(\lambda)\}$. Since the
observables depend only on the densities by assumption, the generic
Bethe Ansatz eigenstate scaling to $\{\rho_{n}(\lambda)\}$ will be
denoted $|\{\rho_{n}(\lambda)\}\rangle$, omitting any further microscopic
labels.

The $n$-holes are defined as positions satisfying the $n$-string
quantization relation following from the Bethe equations, but absent
from the wave function. In the thermodynamic limit the density of
$n$-holes $\rho_{n}^{\text{h}}(\lambda)$ can be defined analogously
to $\rho_{n}(\lambda)$. As a consequence of the Bethe equations,
the densities $\rho_{n}(\lambda)$ and $\rho_{n}^{\text{h}}(\lambda)$
are constrained by the Bethe–Takahashi equations \cite{9780511524332}:

\begin{equation}
a_{n}(\lambda)=\rho_{n}(\lambda)+\rho_{n}^{h}(\lambda)+\sum_{m=1}^{\infty}\int_{-\frac{\pi}{2}}^{\frac{\pi}{2}}d\lambda'T_{nm}(\lambda-\lambda')\rho_{m}(\lambda)\,,\label{eq:BetheTakahashiEquations}
\end{equation}
where 
\begin{equation}
a_{n}(\lambda)=\frac{1}{2\pi}i\frac{d}{d\lambda}\log\left(\frac{\sin(\lambda+in\eta/2)}{\sin(\lambda-in\eta/2)}\right)=\frac{1}{\pi}\frac{\sinh(n\eta)}{\cosh(n\eta)-\cos(2\lambda)}\qquad n\ge1\,,\label{eq:a_nDefinition}
\end{equation}
and
\begin{equation}
T_{nm}(x)=\begin{cases}
a_{|n-m|}(\lambda)+2a_{|n-m|+2}(\lambda)+...+2a_{n+m-2}(\lambda)+a_{n+m}(\lambda)\,, & \textrm{if }m\ne n\\
2a_{2}(\lambda)+2a_{4}(\lambda)+...+2a_{2n-2}(\lambda)+a_{2n}(\lambda)\,, & \textrm{if }m=n\,.
\end{cases}
\end{equation}
The equations (\ref{eq:BetheTakahashiEquations}) can be easily transformed
into a partially decoupled form \cite{9780511524332}:

\begin{equation}
\rho_{n}(\lambda)=\frac{1}{1+\eta_{n}(\lambda)}\left(s(\lambda)\delta_{n,1}+\left[s\star\left(\eta_{n}\rho_{n}+\eta_{n-1}\rho_{n-1}\right)\right](\lambda)\right)\,,\label{eq:BetheTakahashiDecoupled}
\end{equation}
where 
\begin{equation}
\eta_{n}(\lambda)=\frac{\rho_{n}^{\textrm{h}}(\lambda)}{\rho(\lambda)}\,.\label{eq:etadef}
\end{equation}
Note that (\ref{eq:BetheTakahashiEquations}) determine $\{\rho_{n}^{\text{h}}(\lambda)$\}
uniquely if $\{\rho_{n}(\lambda)$\} is given, therefore we will write
functionals of $\{\rho_{n}^{\text{h}}(\lambda)$\} as functionals
$\{\rho_{n}(\lambda)$\} for the sake of brevity. 

The density of particles is given by
\begin{equation}
\mathcal{M}[\{\rho_{n}(\lambda)\}]=\sum_{n=1}^{\infty}\int_{-\pi/2}^{\pi/2}d\lambda n\rho_{n}(\lambda)=\frac{M}{L}\,.\label{eq:densityOfParticles}
\end{equation}
This functional, when restricted to Bethe Ansatz string densities
satisfying (\ref{eq:BetheTakahashiEquations}), can take values from
the interval $[0,\frac{1}{2}]$ and is simply related to the magnetization
per site, with zero magnetization corresponding to $\mathcal{M}=1/2$.

The \emph{XXZ }chain has infinitely many local conserved charges which
are the logarithmic derivatives of the \emph{XXZ} transfer matrix
\cite{9780511628832}:
\begin{equation}
Q_{k}=\left(i\frac{d}{d\lambda}\right)^{k-1}\log T^{\mathit{XXZ}}(\lambda)\big|_{\lambda=0}\,,\label{eq:XXZcharges}
\end{equation}
with $Q_{2}$ proportional to the Hamiltonian. The locality of these
charges means that they are all given as sums over the chain of terms
containing a limited number of spin operators acting on adjacent sites
\cite{1994MPLA....9.2197G}. Their expectation values in Bethe Ansatz
eigenstates can be constructed using the algebraic Bethe Ansatz \cite{9780511628832}.
In the thermodynamic limit, the expectation values of conserved charges
per site in a particular state $|\{\rho_{n}(\lambda)\}\rangle$ are
obtained as a sum of integrals of the string densities with appropriate
kernel functions $q_{n}^{(j)}(\lambda)$: 
\begin{equation}
\langle\{\rho_{n}(\lambda)\}|Q_{k}|\{\rho_{n}(\lambda)\}\rangle=\sum_{n=1}^{\infty}\int_{-\pi/2}^{\pi/2}d\lambda\rho_{n}(\lambda)q_{n}^{(k)}(\lambda)\,,\label{eq:chargesIntegralFormula}
\end{equation}
with

\begin{equation}
\begin{aligned}q_{n}^{(k)}(\lambda) & =-2\pi\left(i\frac{d}{d\lambda}\right)^{k-2}a_{n}(\lambda)\qquad k\ge2\,.\end{aligned}
\end{equation}
It is important to note that a one-to-one correspondence between $\rho_{1}^{\text{h}}(\lambda)$
and the expectation values of conserved charges was derived in \cite{Wouters2014,Brockmann2014b}.
In our conventions, the relation reads
\begin{equation}
\begin{aligned}\langle\{\rho_{n}(\lambda)\}|Q_{j}|\{\rho_{n}(\lambda)\}\rangle & =\sum_{m=-\infty}^{\infty}\frac{\tilde{\rho}_{1}^{\text{h}}(m)-e^{-|m|\eta}}{2\cosh(m\eta)}\,(2mi)^{j-2}\,,\\
\tilde{\rho}_{n}(m) & =\int_{\pi/2}^{\pi/2}d\lambda\rho_{n}(\lambda)e^{im\lambda}\,,\\
\rho_{n}(m) & =\frac{1}{\pi}\sum_{m=-\infty}^{\infty}\tilde{\rho}_{n}(k)e^{-im\lambda}\,.
\end{aligned}
\end{equation}
Following \cite{Fagotti2013}, it is useful to define a generating
function as 
\begin{equation}
G(\lambda)=\sum_{j=0}^{\infty}\frac{\lambda^{j}}{j!}\langle\{\rho_{n}(\lambda)\}|Q_{j+1}|\{\rho_{n}(\lambda)\}\rangle\,,\label{eq:ChargeGenFunDef}
\end{equation}
which is in a one-to-one relationship with $\rho_{1}^{\text{h}}(\lambda)$
\cite{Wouters2014,Brockmann2014b}:
\begin{equation}
G(\lambda)=[s\star(\rho_{1}^{\text{h}}-a_{n})](\lambda)\,,\label{eq:generatingFunctionRho1h}
\end{equation}
with 
\begin{equation}
s(\lambda)=\frac{1}{2\pi}\left(1+2\sum_{k=1}^{\infty}\frac{\cos2k\lambda}{\cosh k\eta}\right)\,.\label{eq:sFunctionDefinition}
\end{equation}

\section{Bethe Ansatz for quantum quenches in the XXZ spin chain\label{sec:Bethe-Ansatz-for-QQ-in-XXZ}}

In general, quantum quenches are described by the following protocol: 
\begin{enumerate}
\item The system is initially prepared in a ground state $|\Psi_{0}\rangle$
of a local Hamiltonian $H_{0}$. 
\item At $t=0$ the Hamiltonian is suddenly changed, and from then on, the
system evolves in time according to the new or \emph{post-quench}
Hamiltonian $H$. 
\item After a suitably long time, the system is expected to relax into a
steady state.
\end{enumerate}
We only consider translation invariant (global) quantum quenches when
both $H_{0}$ and $H$ are translationally invariant and the quench
is realized by changing one or more coupling constant in the Hamiltonian
at $t=0$. The post-quench Hamiltonian $H$ is the \emph{XXZ} Hamiltonian
(\ref{eq:XXZHamiltonian}), while the initial states are certain translationally
invariant product states $|\Psi_{0}^{\text{\ensuremath{\gamma}}}\rangle$
of the form
\begin{equation}
\begin{aligned}|\Psi_{0}^{\text{\ensuremath{\gamma}}}\rangle & =\frac{1+\hat{T}}{\sqrt{2}}|\psi_{0}^{\gamma}\rangle\\
|\psi_{0}^{\gamma}\rangle & =\left[\otimes^{L/2}\left(\frac{|\uparrow\downarrow\rangle-\gamma|\downarrow\uparrow\rangle}{\sqrt{1+\gamma^{2}}}\right)\right]\,,
\end{aligned}
\label{eq:initialStates}
\end{equation}
where $\hat{T}$ is the one site translation operator and $\gamma$
is a constant determining the specific initial state. In this study,
three different values of $\gamma$ are considered:
\begin{enumerate}
\item $\gamma=0$ is the translationally invariant Néel state, which is
a ground state of the $\Delta\rightarrow\infty$ limit of the\emph{
XXZ }Hamiltonian, and will also be denoted by $|\Psi_{0}^{\text{N}}\rangle=|\Psi_{0}^{0}\rangle$.
\item $\gamma=1$ is translationally invariant Majumdar-Ghosh dimer product
state, which is a ground state of the Majumdar-Ghosh Hamiltonian \cite{1969JMP....10.1388M},
and will also be denoted by $|\Psi_{0}^{\text{D}}\rangle=|\Psi_{0}^{1}\rangle$. 
\item $\gamma=q$, where $q+1/q=\Delta$, is the translationally invariant
$q$-deformed dimer product state, a ground state of the $q$-deformed
Majumdar-Ghosh Hamiltonian \cite{BATCHELOR1994}, which will be alternatively
denoted by $|\Psi_{0}^{q\text{D}}\rangle=|\Psi_{0}^{q}\rangle$.
\end{enumerate}
Although the quenches start from the translationally invariant states
defined above, for later convenience their non-translationally invariant
counterparts $|\psi_{0}^{0}\rangle$, $|\psi_{0}^{1}\rangle$ and
$|\psi_{0}^{q}\rangle$ will be denoted by $|\text{N}\rangle$, $|\text{D}\rangle$
and $|q\text{D}\rangle$, respectively.

\subsection{The diagonal ensemble\label{sub:The-diagonal-ensemble}}

The goal is to compute the infinite time average of the expectation
value of local observables $\mathcal{O}$ in the thermodynamic limit
\begin{equation}
\langle\mathcal{O}\rangle=\lim_{T\rightarrow\infty}\frac{1}{T}\int\limits _{0}^{T}dt\langle\Psi_{0}|e^{iHt}\mathcal{O}e^{-iHt}|\Psi_{0}\rangle\qquad(\TDL)\,.\label{eq:timeAverage}
\end{equation}
For well-behaved initial states, the average (\ref{eq:timeAverage})
can be computed as an ensemble average of the so-called \emph{diagonal
ensemble}. Expanding the expectation value $\langle\Psi_{0}|e^{iHt}\mathcal{O}e^{-iHt}|\Psi_{0}\rangle$
over the eigenstates of the post-quench Hamiltonian as
\begin{equation}
\langle\Psi_{0}|e^{iHt}\mathcal{O}e^{-iHt}|\Psi_{0}\rangle=\sum_{\alpha'}\sum_{\alpha}e^{-i(E_{\alpha}-E_{\alpha'})t}\langle\Psi_{0}|\alpha'\rangle\langle\alpha'|\mathcal{O}|\alpha\rangle\langle\alpha|\Psi_{0}\rangle\,,\label{eq:doubleSum}
\end{equation}
where $E_{\alpha}$ is the energy of the Hamiltonian eigenstate $|\alpha\rangle$,
and substituting (\ref{eq:doubleSum}) into (\ref{eq:timeAverage}):

\begin{equation}
\langle\mathcal{O}\rangle=\lim_{T\rightarrow\infty}\frac{1}{T}\int\limits _{0}^{T}dt\sum_{\alpha'}\sum_{\alpha}e^{-i(E_{\alpha}-E_{\alpha'})t}\langle\Psi_{0}|\alpha'\rangle\langle\alpha'|\mathcal{O}|\alpha\rangle\langle\alpha|\Psi_{0}\rangle\qquad(\TDL)\,.\label{eq:doubleSumTimeAverage}
\end{equation}
For non-degenerate systems and/or suitably generic starting states,
this expression simplifies to a single sum when taking the limits
in the given order, since the off-diagonal terms contain rapidly oscillating
exponentials that cancel out. The remaining terms give an average
over the so called \emph{diagonal ensemble:} 
\begin{equation}
\langle\mathcal{O}\rangle=\sum|\langle\Psi_{0}|\alpha\rangle|^{2}\langle\alpha|\mathcal{O}|\alpha\rangle\qquad(\TDL)\,.\label{eq:diagonalEnsemble}
\end{equation}
The validity of the diagonal ensemble (\ref{eq:diagonalEnsemble})
is the underlying assumption of both the generalized Gibbs ensemble
hypothesis and the quench action formalism, which are introduced in
the remainder of this section.

\subsection{Remarks on the role of translational invariance}

The translational invariance of the chosen initial states $|\Psi_{0}^{\gamma}\rangle$
is important because by (\ref{eq:HamiltonianTranslationCommutator})
it assures the translational invariance of the steady state and thus
the validity of the diagonal ensemble (\ref{eq:diagonalEnsemble}).
It is clear that if the post-quench steady values are not translationally
invariant, then they cannot be described by the diagonal ensemble
(\ref{eq:diagonalEnsemble}).

Our initial states have the form (\ref{eq:initialStates})

\[
|\Psi_{0}^{\text{\ensuremath{\gamma}}}\rangle=\frac{1+\hat{T}}{\sqrt{2}}|\psi_{0}^{\gamma}\rangle
\]
where the states $|\psi_{0}^{\gamma}\rangle$ are invariant under
$\hat{T}^{2}$ and
\begin{equation}
[H_{\mathrm{XXZ}}(\Delta),\hat{T]}=0\,,\label{eq:HamiltonianTranslationCommutator}
\end{equation}
The states $|\psi_{0}^{\gamma}\rangle$ have non-zero overlaps only
with Hamiltonian eigenstates $|\alpha\rangle$ that satisfy
\begin{equation}
\hat{T}|\alpha\rangle=\pm|\alpha\rangle\,,\label{eq:eigenstateTranslationEigenvalue}
\end{equation}
which ensures that the diagonal terms of the double sum in (\ref{eq:doubleSumTimeAverage})
are translationally invariant because of (\ref{eq:HamiltonianTranslationCommutator}-\ref{eq:eigenstateTranslationEigenvalue}).
Therefore whenever translational invariance is broken, the off-diagonal
terms cannot cancel. In addition, if the diagonal ensemble is not
valid, then neither the GGE nor the quench action method can describe
for the steady state, as they both assume the validity of the diagonal
ensemble.

On the other hand, the general validity of the diagonal ensemble is
not clear for translational invariance breaking initial states; in
particular, it is an open problem whether translational invariance
of observables is restored after quenches from the state $|\text{D}\rangle$
\cite{Fagotti2014,Fagotti2014a}. Nevertheless, for translationally
invariant initial states $|\Psi_{0}^{\gamma}\rangle$ then the post-quench
steady state will be translationally invariant, since the post-quench
Hamiltonian preserves translational invariance.

\subsection{GGE and GTBA}

\subsubsection{The generalized Gibbs ensemble}

The idea of the generalized Gibbs ensemble \cite{Rigol2007} is to
include all the relevant conserved charges $Q_{j}$ in the statistical
operator with appropriate Lagrange multipliers $\beta_{j}$ 

\begin{equation}
\rho_{\text{GGE}}=\frac{1}{Z_{\text{GGE}}}\exp\!\left(\sum\limits _{k=1}^{\infty}\beta_{k}Q_{k}\right)\qquad Z_{\text{GGE}}=\Tr\,\exp\!\left(\sum\limits _{k=1}^{\infty}\beta_{k}Q_{k}\right)\,,\label{eq:GGE}
\end{equation}
in order to set the ensemble averages of the conserved charges to
their initial state expectation value 
\[
\langle\Psi_{0}|Q_{k}|\Psi_{0}\rangle=\Tr\,\rho_{\text{GGE}}Q_{k}\,.
\]
According to the hypothesis corresponding to the generalized Gibbs
ensemble, the expectation value of observables in the post-quench
relaxed state{]} may be expressed using $\rho_{GGE}$:
\[
\langle\mathcal{O}\rangle_{\text{GGE}}=\Tr\rho_{GGE}\mathcal{O}\,.
\]
The GGE is the ensemble which follows from conditional entropy maximization
while keeping the expectation values of the charges fixed to their
pre-quench values.

What are the relevant charges to include in the GGE statistical operator?
It is clear that for quenches starting from a pure state the full
quantum state of the system remains a pure state for all times. As
a result, the GGE can only be valid for some specific class of observables,
which we call \emph{relevant}. Due to the fact that both the pre-quench
and post-quench Hamiltonians are local in spatial sense, here we choose
these observables as local correlations of spins of the form $\sigma_{j}\sigma_{j+l}$,
where the distance $l$ remains finite while $L\rightarrow\infty$
in the thermodynamic limit. The relevant conserved charges are then
expected to be the local charges $Q_{k}$, since the $k$th charge
can be expressed as a sum over terms containing products of at most
$k$ adjacent spin operators. This reasoning leads to the definition
of the GGE by including only local conserved charges and restricting
the class of relevant observables to local ones, as emphasized e.g.
in \cite{Fagotti2014}. We shall return to the role of locality in
the discussion.

\subsubsection{GGE and GTBA for the XXZ chain \label{sub:GTBA}}

For the XXZ chain the GGE predictions can be computed using two different
methods: the quantum transfer matrix (QTM) method \cite{1993ZPhyB..91..507K,Klumper2004}
and a generalized thermodynamic Bethe Ansatz (GTBA) \cite{2012JPhA...45y5001M}. 

The QTM method for obtaining the mean values of local correlators
$\sigma_{j}\sigma_{j+l}$ for a truncated GGE with the statistical
operator
\begin{equation}
\rho_{\text{TGGE}}=\frac{1}{Z_{\text{TGGE}}}\exp\!\left(\sum\limits _{k=1}^{k_{max}}\beta_{k}Q_{k}\right)\hspace{1em}Z_{\text{TGGE}}=\Tr\,\exp\!\left(\sum\limits _{k=1}^{k_{max}}\beta_{k}Q_{k}\right),
\end{equation}
was initially developed in  \cite{Pozsgay2013b}, and involved only
the first $k_{max}=12$ conserved charges. The reason why this truncation
works is that the charges with the lowest indices are also the most
local ones, and correlations over a distance $l$ are considered to
be insensitive to the value of charges with $k>l$. Shortly thereafter
the QTM formalism for the full GGE (\ref{eq:GGE}) was constructed
in a large $\Delta$ limit \cite{Fagotti2013}, and then for arbitrary
$\Delta>0$ \cite{Fagotti2014}. The numerical results of  \cite{Pozsgay2013b}
obtained through the truncated GGE approximate the full GGE results
quite well, so the GGE is truncatable in the sense that keeping the
first few local charges gives a good approximation of the full result,
which is systematically improved by increasing $k_{max}$.

Another possibility of computing the GGE predictions for correlations
is using the TBA method for the GGE which was derived in  \cite{Wouters2014}.
Using the entropy maximization principle with the condition 
\begin{equation}
\langle\{\rho_{n}(\lambda)\}|Q_{k}|\{\rho_{n}(\lambda)\}\rangle=\langle\Psi_{0}|Q_{k}|\Psi_{0}\rangle\,,\label{eq:charges_fixed}
\end{equation}
a set of generalized TBA (GTBA) equations can be derived, which determine
the Bethe Ansatz string densities $\{\rho_{n}(\lambda)$\} which maximizing
the entropy (\ref{eq:entropyDefinition}) under the conditions (\ref{eq:charges_fixed}).
The GTBA equations \cite{Wouters2014,Brockmann2014b}
\begin{equation}
\ln\eta_{n}(\lambda)=-\delta_{n,1}\sum_{k=0}^{\infty}\beta_{k+2}\left(\frac{d}{d\lambda}\right)^{k}s(\lambda)+\left[s\star\left(\ln(1+\eta_{n-1})+\ln(1+\eta_{n+1})\right)\right](\lambda)\label{eq:GGEGaudinTakahashiEquation}
\end{equation}
are to be solved for the functions $\eta_{n}$ defined in (\ref{eq:etadef}),
after which the Bethe-Takahashi equations (\ref{eq:BetheTakahashiDecoupled})
can be used to obtain the string densities. 

However, (\ref{eq:GGEGaudinTakahashiEquation}) contains infinitely
many unknown Lagrange multipliers that must be fixed using (\ref{eq:charges_fixed}).
It was found in \cite{Wouters2014} that the system (\ref{eq:GGEGaudinTakahashiEquation})
can be solved by using the relation (\ref{eq:generatingFunctionRho1h})
between the generating function of the charges and $\rho_{1}^{\textrm{h}}(\lambda)$,
thus avoiding the determination of Lagrange multipliers. Provided
$\rho_{1}^{\textrm{h}}(\lambda)$ is known, then the equations for
$\rho_{n}(\lambda)$ can be recast into the following form \cite{2014JSMTE..09..026P,Brockmann2014b}:
\begin{equation}
\begin{aligned}\log\eta_{2}(\lambda) & =\mbox{s\ensuremath{\star\left[\log\left(\frac{s+s\star\eta_{2}\rho_{2}}{s+s\star\eta_{2}\rho_{2}-\rho_{1}^{\textrm{h}}}\right)+\log\left(1+\eta_{3}\right)\right]}}(\lambda)\\
\rho_{2}(\lambda) & =\frac{1}{1+\eta_{2}(\lambda)}\left[s\star\left(\rho_{1}^{\textrm{h}}+\eta_{n-1}\rho_{n-1}\right)\right](\lambda)\\
\ln\eta_{n}(\lambda) & =\left[s\star\left(\ln(1+\eta_{n-1})+\ln(1+\eta_{n+1})\right)\right](\lambda)\qquad n\ge3\\
\rho_{n}(\lambda) & =\frac{1}{1+\eta_{n}(\lambda)}\left[s\star\left(\eta_{n}\rho_{n}+\eta_{n-1}\rho_{n-1}\right)\right](\lambda)\qquad n\ge3\:,
\end{aligned}
\label{eq:GGETBAEquations}
\end{equation}
for $\eta_{n}(\lambda)$ and $\rho_{n}(\lambda)$, then using (\ref{eq:GGEGaudinTakahashiEquation})
and (\ref{eq:BetheTakahashiDecoupled}) with $n=1$ to obtain $\rho_{1}(\lambda)$.
In the thermodynamic limit, these string densities determine the expectation
value of every relevant observable; in particular, the results of
 \cite{Mestyan2014} allow to compute arbitrary short-range correlations
from $\{\rho_{n}(\lambda)$\}.

For the particular initial states considered in this work, the generating
functions of conserved charges are
\begin{eqnarray}
\mbox{Néel} & : & G_{N}(\lambda)=-\frac{\sinh2\eta}{\cosh2\eta+1-2\cos2\lambda}\nonumber \\
\mbox{Dimer} & : & G_{D}(\lambda)=-\sinh\eta\frac{4\cos2\lambda(\sinh^{2}\eta-\cosh\eta)+\cosh\eta+2\cosh2\eta+3\cosh3\eta-2}{4\left(\cosh2\eta-\cos2\lambda\right)^{2}}\nonumber \\
\mbox{q-dimer} & : & G_{qD}(\lambda)=\tanh\eta\frac{2\cos2\lambda-\cosh2\eta-\cosh4\eta}{2\left(\cosh2\eta-\cos2\lambda\right)^{2}}\label{eq:ChargeGenFunc}
\end{eqnarray}
The first two results were explicitly computed in \cite{Fagotti2014},
while the third one can be obtained straightforwardly using the formalism
developed there.

\subsection{The quench action approach}

The quench action approach was introduced as a way of computing long-time
expectation values of local observables after quenches to Bethe Ansatz
solvable systems \cite{2013PhRvL.110y7203C}. In this study we only
consider infinite time expectation values, and summarize the idea
of the quench action with the corresponding simplifications. The first
step is to replace the sums in (\ref{eq:diagonalEnsemble}) by a functional
integral over Bethe root densities:
\[
\sum_{\alpha}\rightarrow\int\prod_{n=1}^{\infty}D\rho_{n}(\lambda)e^{Ls[\{\rho_{n}(\lambda)\}]}\,,
\]
where the exponential of the entropy $Ls[\{\rho_{n}(\lambda)\}]$
is the number of Bethe states scaling to the set of densities $\{\rho_{n}(\lambda)\}$.
The expression (\ref{eq:diagonalEnsemble}) then takes the form
\begin{equation}
\langle\mathcal{O}\rangle=\int\prod_{n=1}^{\infty}D\rho_{n}(\lambda)e^{-L\left(-\frac{2}{L}\mbox{Re}\ln\langle\Psi_{0}|\{\rho_{n}(\lambda)\}\rangle-s[\{\rho_{n}(\lambda)\}]\right)}\langle\{\rho_{n}(\lambda)\}|\mathcal{O}|\{\rho_{n}(\lambda)\}\rangle\,.\label{eq:diagonalEnsembleBA}
\end{equation}
In the thermodynamic limit the functional integral can be evaluated
exactly using saddle point analysis. The saddle point string densities
$\{\rho_{n}^{*}(\lambda)\}$ minimize the quench action functional
\begin{equation}
\mathcal{S}[\{\rho_{n}(\lambda)\}]=-\frac{2}{L}\mbox{Re}\ln\langle\Psi_{0}|\{\rho_{n}(\lambda)\}\rangle-s[\{\rho_{n}(\lambda)\}]\,,\label{eq:quenchAction}
\end{equation}
with the condition that the Bethe-Takahashi equations (\ref{eq:BetheTakahashiEquations})
hold. The quench action is analogous to the free energy functional
appearing in the thermal thermodynamic Bethe Ansatz \cite{9780511524332}.
The first term, which parallels the energy in the context of a thermal
ensemble, competes with the second entropic term. When evaluated at
the saddle point the quench action gives the norm of the initial state:
\begin{equation}
L\mathcal{S}[\{\rho_{n}^{*}(\lambda)\}]=-\ln\langle\Psi_{0}|\Psi_{0}\rangle=0\:,\label{eq:OverlapSumRule}
\end{equation}
which is a sum rule that can be used to check whether the relevant
saddle point was found. 

In terms of the saddle point string densities that minimize the quench
action, the diagonal ensemble average (\ref{eq:diagonalEnsembleBA})
can be expressed as 
\begin{equation}
\langle\mathcal{O}\rangle=\langle\{\rho_{n}^{*}(\lambda)\}|\mathcal{O}|\{\rho_{n}^{*}(\lambda)\}\rangle.\label{eq:saddlePointExpValue}
\end{equation}
The following two sections discuss the computation of (\ref{eq:saddlePointExpValue})
for certain quenches of the \emph{XXZ} spin chain. The variational
analysis yielding the steady state $\{\rho_{n}^{*}(\lambda)\}$ is
treated in Section \ref{sec:steadyStateDensities}, and Section \ref{sec:Computing-correlation-functions}
deals with the calculation of expectation values in the Bethe Ansatz
eigenstate characterized by $\{\rho_{n}^{*}(\lambda)\}$.

\section{Computing the steady state of \emph{XXZ} using the quench action\label{sec:steadyStateDensities}}

\subsection{Overlaps of the initial states with Bethe Ansatz eigenstates}

Since $H_{\textit{XXZ}}$ commutes with component $z$ of the total
spin, the overlap of the above initial states is nonzero only with
Bethe states that have $\mathcal{M}[\{\rho_{n}(\lambda)\}]=\frac{1}{2}$.
As shown below, for such states the first term of (\ref{eq:quenchAction})
can be written in the integral form in the the thermodynamic limit

\begin{equation}
\begin{aligned}-2\,\mbox{Re}\ln\langle\Psi_{0}^{\gamma}|\{\rho_{n}(\lambda)\}\rangle & =\sum_{n=1}^{\infty}\int_{-\pi/2}^{\pi/2}d\lambda\rho_{n}(\lambda)g_{n}^{(\gamma)}(\lambda)\\
g_{n}^{(\gamma)}(\lambda) & =\sum_{j=1}^{n}g_{1}^{(\gamma)}\left(\lambda+\frac{i\eta}{2}(n+1-2j)\right)\,,
\end{aligned}
\label{eq:overlapIntegralForm}
\end{equation}
where the one-string kernel function $g_{1}^{(\gamma)}(\lambda)$
corresponds to the particular initial state chosen. This integral
form is very convenient in the sense that the variational equations
for minimizing $\mathcal{S}[\{\rho_{n}(\lambda)\}]$ are analogous
to the variational equations for minimizing the free energy in the
context of the thermodynamic Bethe Ansatz \cite{9780511524332}. 

However, the integral formula (\ref{eq:overlapIntegralForm}) yields
finite values also for Bethe states with nonzero overall magnetization,
thus its naive use in the variational problem leads to spurious results.
The Bethe states with $\mathcal{M}[\{\rho_{n}(\lambda)\}]\neq\frac{1}{2}$
can be excluded explicitly from the set of possible solutions by varying
\begin{equation}
\mathcal{\tilde{\mathcal{S}}}[\{\rho_{n}(\lambda)\}]=\sum_{n=1}^{\infty}\int_{-\pi/2}^{\pi/2}d\lambda\rho_{n}(\lambda)g_{n}^{(\gamma)}(\lambda)-\mu\mathcal{M}[\{\rho_{n}(\lambda)\}]-s[\{\rho_{n}(\lambda)\}]\,,\label{eq:quenchActionLagrange}
\end{equation}
where $\mu$ is a Lagrange multiplier used to set $\mathcal{M}[\{\rho_{n}^{*}(\lambda)\}]=\frac{1}{2}$
\cite{Pozsgay2014,Wouters2014}. We remark here that taking $\mu\rightarrow\infty$
limits the possible solutions to the shell $\mathcal{M}[\{\rho_{n}^{*}(\lambda)\}]=\frac{1}{2}$,
since any thermodynamic Bethe Ansatz state has $\mathcal{M}[\{\rho_{n}(\lambda)\}]\leq\frac{1}{2}$.
In  \cite{Brockmann2014b}, it was proved that $\mu$ \emph{has to
be} infinity by using the asymptotics of the variational equations
for $n\rightarrow\infty$. For a numerical solution of the variational
problem, it only matters to choose $\mu$ large enough to impose $\mathcal{M}[\{\rho_{n}(\lambda)\}]=\frac{1}{2}$
within the accuracy of the numerical solution.

Before moving on to the variational equations for minimizing $\mathcal{\tilde{\mathcal{S}}}[\{\rho_{n}(\lambda)\}]$,
we summarize the derivation of the one-string kernels $g_{1}^{(\gamma)}(\lambda)$
for different initial states.

\subsubsection{The Néel state}

It was shown in  \cite{Brockmann2014} that only the eigenstates $|\{\pm\lambda_{j}\}_{j=1}^{M/2}\rangle$
containing pairs of rapidities have non-zero overlaps with $|\Psi_{0}^{\text{N}}\rangle$.
The logarithmic overlap of the finite volume translationally invariant
Néel state with Bethe Ansatz eigenstates of the form 
\begin{equation}
|\{\pm\lambda_{j}\}_{j=1}^{M/2}\rangle\quad M=L/2\,,
\end{equation}
i.e. paired states having zero total magnetization, was derived in
Refs. \cite{2014JPhA...47n5003B,Brockmann2014c}. The result is
\begin{equation}
\ln\frac{\langle\Psi_{0}^{\text{N}}|\{\pm\lambda_{j}\}_{j=1}^{M/2}\rangle}{\sqrt{\langle\{\pm\lambda_{j}\}_{j=1}^{M/2}|\{\pm\lambda_{j}\}_{j=1}^{M/2}\rangle}}=\sum_{j=1}^{M/2}\ln\left(\frac{\sqrt{\tan(\lambda_{j}+i\eta/2)\tan(\lambda_{j}-i\eta/2)}}{2\sin(2\lambda_{j})}\right)+\frac{1}{2}\ln\frac{2\det_{L/4}G_{jk}^{+}}{\det_{L/4}G_{jk}^{-}}\,,\label{eq:NeelOverlapFiniteVolume}
\end{equation}
where 
\begin{eqnarray}
G_{jk}^{\pm} & = & \delta_{jk}\left(NK_{\eta/2}(\lambda_{j})-\sum_{l=1}^{L/4}K_{\eta}^{\pm}(\lambda_{j},\lambda_{l})\right)+K_{\eta}^{\pm}(\lambda_{j},\lambda_{k})\nonumber \\
K^{\pm}(\lambda,\mu) & = & K(\lambda-\mu)\pm K(\lambda+\mu)\nonumber \\
K(\lambda) & = & \frac{\sinh(2\eta)}{\sinh(\lambda+i\eta)\sinh(\lambda-i\eta)}\,.
\end{eqnarray}
The second term of (\ref{eq:NeelOverlapFiniteVolume}) scales as $\mathcal{O}(1)$
and therefore it is negligible in the thermodynamic limit \cite{Brockmann2014b},
while the thermodynamic limit of the first term in (\ref{eq:NeelOverlapFiniteVolume})
yields the one-string kernel of (\ref{eq:overlapIntegralForm}) corresponding
to the initial state $|\Psi_{0}^{\text{N}}\rangle$:
\begin{equation}
g_{1}^{\text{N}}(\lambda)=-\ln\left(\frac{\tan(\lambda+i\eta/2)\tan(\lambda-i\eta/2)}{4\sin^{2}(2\lambda)}\right)\,.\label{eq:NeelOverlapKernel}
\end{equation}
To check the validity of the formulas (\ref{eq:overlapIntegralForm})
together with the kernel (\ref{eq:NeelOverlapKernel}), we computed
the quantity $2\,\mbox{Re}\ln\langle\Psi_{0}^{\text{N}}|\Psi_{\Delta}^{\text{GS}}\rangle$
for different values of $\Delta>1$, where $|\Psi_{\Delta}^{\text{GS}}\rangle$
is the ground state of $H_{XXZ}(\Delta)$. The state $|\Psi_{\Delta}^{\text{GS}}\rangle$
consists of 1-strings only, and its root density is given by an inverse
Fourier transform formula \cite{9780511524332}:
\begin{equation}
\rho_{1,\Delta}^{\text{GS}}(\lambda)=\frac{1}{\pi}\sum_{k=-\infty}^{\infty}\frac{1}{2\cosh k\eta}e^{i2k\lambda},\hspace{1em}\rho_{n,\Delta}^{\text{GS }}(\lambda)=0\hspace{1em}(n>1)\,.\label{eq:GS_rootdensity}
\end{equation}
The $\TDL$ values of 
\[
2\,\mbox{Re}\ln\langle\Psi_{0}^{\text{N}}|\Psi_{\Delta}^{\text{GS}}\rangle\,,
\]
obtained numerically from (\ref{eq:overlapIntegralForm},\ref{eq:GS_rootdensity})
are in exact match with the values presented in  \cite{Pozsgay2013},
computed independently by taking the $t\rightarrow i\infty$ limit
of the Loschmidt echo defined by 
\[
\langle\Psi_{0}^{\text{N}}|e^{-iH_{\textit{XXZ}}(\Delta)t}|\Psi_{0}^{\text{N}}\rangle\,.
\]

\subsubsection{\label{sub:The-dimer-state} The dimer state }

The overlaps of the dimer state can be easily computed provided that
the overlaps of the Néel state are known. The following relation holds
between the overlaps of the Dimer state and the overlaps of the Néel
state with general zero magnetization ($M=L/2$) Bethe states: 
\begin{equation}
\langle\text{D}|\{\lambda_{j}\}_{j=1}^{M}\rangle=\langle\text{N}|\{\lambda_{j}\}_{j=1}^{M}\rangle\prod_{j=1}^{M}\frac{1}{\sqrt{2}}\left(1-\frac{\sin(\lambda_{j}-i\eta/2)}{\sin(\lambda_{j}+i\eta/2)}\right)\,.\label{eq:NeelDimerRelation}
\end{equation}
This relation, first published without derivation in \cite{Pozsgay2014a},
follows from the formula 
\begin{equation}
|\text{D}\rangle=\prod_{k=1}^{M}\left(\frac{1-S_{2k-1}^{-}S_{2k}^{+}}{\sqrt{2}}\right)|\text{N}\rangle\,,
\end{equation}
and the form (\ref{eq:basisExpansion}-\ref{eq:BetheWaveFunction})
of the Bethe Ansatz wave functions:
\begin{equation}
\begin{aligned}\langle\text{D}|\{\lambda_{j}\}_{j=1}^{M}\rangle & =\langle\text{N}|\prod_{k=1}^{M}\left(\frac{1-S_{2k-1}^{+}S_{2k}^{-}}{\sqrt{2}}\right)|\{\lambda_{j}\}_{j=1}^{M}\rangle\\
 & =\langle\text{N}|\text{N}\rangle\sum_{\pi\in S_{M}}I(\pi)\prod_{k=1}^{M}\frac{1}{\sqrt{2}}\left(1-\frac{\sin(\lambda_{\pi(k)}+i\eta/2)}{\sin(\lambda_{\pi(k)}-i\eta/2)}\right)\\
 & \qquad\times\prod_{j=1}^{M}\left(\frac{\sin(\lambda_{j}+i\eta/2)}{\sin(\lambda_{j}-i\eta/2)}\right)^{n_{\pi(j)}}\left(\prod_{j<k}...\right)\\
 & =\left[\langle\text{N}|\text{N}\rangle\sum_{\pi\in S_{M}}I(\pi)\prod_{j=1}^{M}\left(\frac{\sin(\lambda_{j}+i\eta/2)}{\sin(\lambda_{j}-i\eta/2)}\right)^{n_{\pi(j)}}\left(\prod_{j<k}...\right)\right]\\
 & \qquad\times\prod_{k=1}^{M}\frac{1}{\sqrt{2}}\left(1-\left(\frac{\sin(\lambda_{k}+i\eta/2)}{\sin(\lambda_{k}-i\eta/2)}\right)\right)\\
 & =\langle\text{N}|\{\lambda_{j}\}_{j=1}^{M}\rangle\prod_{k=1}^{M}\frac{1}{\sqrt{2}}\left(1-\frac{\sin(\lambda_{k}+i\eta/2)}{\sin(\lambda_{k}-i\eta/2)}\right)\,.
\end{aligned}
\end{equation}
We stress that (\ref{eq:NeelDimerRelation}) is only valid for $M=L/2$. 

A formula similar to (\ref{eq:NeelDimerRelation}) is true for the
translationally invariant states $|\Psi_{0}^{\text{D}}\rangle$ and
$|\Psi_{0}^{\text{N}}\rangle$ since the the Bethe states are eigenstates
of the one site translation operator $\hat{T}$: 
\begin{equation}
\begin{aligned}\langle\Psi_{0}^{\text{D}}|\{\lambda_{j}\}_{j=1}^{M}\rangle & =\langle D|\frac{1+\hat{T}}{\sqrt{2}}|\{\lambda_{j}\}_{j=1}^{M}\rangle=\frac{1}{\sqrt{2}}\left(1+\prod_{j=1}^{M}\left(\frac{\sin(\lambda_{j}+i\eta/2)}{\sin(\lambda_{j}-i\eta/2)}\right)\right)\langle\text{D}|\{\lambda_{j}\}_{j=1}^{M}\rangle\\
 & =\frac{1}{\sqrt{2}}\left(1+\prod_{j=1}^{M}\left(\frac{\sin(\lambda_{j}+i\eta/2)}{\sin(\lambda_{j}-i\eta/2)}\right)\right)\langle\text{N}|\{\lambda_{j}\}_{j=1}^{M}\rangle\\
 & \qquad\times\prod_{j=1}^{M}\frac{1}{\sqrt{2}}\left(1-\frac{\sin(\lambda_{j}-i\eta/2)}{\sin(\lambda_{j}+i\eta/2)}\right)\\
 & =\langle\text{N}|\frac{1+\hat{T}}{\sqrt{2}}|\{\lambda_{j}\}_{j=1}^{M}\rangle\prod_{j=1}^{M}\frac{1}{\sqrt{2}}\left(1-\frac{\sin(\lambda_{j}-i\eta/2)}{\sin(\lambda_{j}+i\eta/2)}\right)\\
 & =\langle\Psi_{0}^{\text{N}}|\{\lambda_{j}\}_{j=1}^{M}\rangle\prod_{j=1}^{M}\frac{1}{\sqrt{2}}\left(1-\frac{\sin(\lambda_{j}-i\eta/2)}{\sin(\lambda_{j}+i\eta/2)}\right)\,.
\end{aligned}
\label{eq:translationalInvariantDimerRelation}
\end{equation}
The logarithmic overlap of the translationally invariant dimer state
is therefore
\begin{equation}
\ln\frac{\langle\Psi_{0}^{\text{D}}|\{\pm\lambda_{j}\}_{j=1}^{M/2}\rangle}{\sqrt{\langle\{\pm\lambda_{j}\}_{j=1}^{M/2}|\{\pm\lambda_{j}\}_{j=1}^{M/2}\rangle}}=\sum_{j=1}^{M/2}\frac{\sinh^{\text{2}}\eta/2\cot\lambda_{j}}{\sqrt{\sin(2\lambda+i\eta)\sin(2\lambda-i\eta)}}+\mathcal{O}(1)\,,
\end{equation}
from which it follows that the logarithmic overlap of $|\Psi_{0}^{\text{D}}\rangle$
with Bethe states $|\{\rho_{n}(\lambda)\}\rangle$ also has the form
(\ref{eq:overlapIntegralForm}) in the thermodynamic limit with the
one-string kernel function being
\begin{equation}
g_{1}^{\text{D}}(\lambda)=-\ln\left(\frac{\sinh^{4}(\eta/2)\cot^{2}(\lambda)}{\sin(2\lambda+i\eta)\sin(2\lambda-i\eta)}\right)\,.
\end{equation}

\subsubsection{The $q$-deformed dimer state}

For the overlaps of the $q$-deformed dimer state a relation similar
to (\ref{eq:NeelDimerRelation}) holds:

\begin{equation}
\langle q\text{D}|\{\lambda_{j}\}_{j=1}^{L/2}\rangle=\langle\text{N}|\{\lambda_{j}\}_{j=1}^{L/2}\rangle\prod_{j=1}^{M}\frac{1}{\sqrt{q+1/q}}\left(q^{-1/2}-q^{1/2}\frac{\sin(\lambda_{j}-i\eta/2)}{\sin(\lambda_{j}+i\eta/2)}\right)\,.\label{eq:NeelqDimerRelation}
\end{equation}
which can be derived similarly to (\ref{eq:NeelDimerRelation}). The
one-string kernel function corresponding to the $q$-deformed dimer
state is thus
\begin{equation}
\begin{aligned}g_{1}^{q\text{D}}(\lambda) & =-\ln\frac{\sinh^{2}\eta}{4\cosh\eta\sin(2\lambda+i\eta)\sin^{2}(2\lambda)\sin(2\lambda-i\eta)}\,.\end{aligned}
\end{equation}

\subsection{Overlap thermodynamic Bethe Ansatz equations}

Using the explicit formulae for the overlaps, it is now possible to
turn the quench action principle into a system of equations for the
string densities characterizing the asymptotic steady state. For the
Néel state, this was obtained prior to our results \cite{Wouters2014,Brockmann2014b};
using our results for the overlaps of the dimer and $q$-dimer state,
we obtain the equations for these initial states as well.

To achieve this, we must consider the variational problem of finding
the string densities $\{\rho_{n}^{*}(\lambda)\}$ that minimize (\ref{eq:quenchActionLagrange})
for the considered quenches of the \emph{XXZ} chain. The integral
formula (\ref{eq:overlapIntegralForm}) can be substituted into first
term of (\ref{eq:quenchActionLagrange}), while the entropy term $s${[}\{$\rho(\lambda)$\}{]}
of (\ref{eq:quenchActionLagrange}) is half of the usual Yang-Yang
entropy,
\begin{equation}
s[\{\rho_{n}(\lambda)\}]=\frac{1}{2}\sum_{n=1}^{\infty}\int_{-\pi/2}^{\pi/2}d\lambda\left[\rho_{n}(\lambda)\ln\left(1+\frac{\rho_{n}^{\text{h}}(\lambda)}{\rho_{n}(\lambda)}\right)+\rho_{n}^{\text{h}}(\lambda)\ln\left(1+\frac{\rho_{n}(\lambda)}{\rho_{n}^{\text{h}}(\lambda)}\right)\right]\,.\label{eq:YangYangEntropyHalf}
\end{equation}
The factor $\frac{1}{2}$ takes into account that only the Bethe Ansatz
eigenstates consisting of rapidity pairs $(\lambda,-\lambda)$ contribute.
Collecting all terms, the functional (\ref{eq:quenchActionLagrange})
takes the form
\begin{equation}
\mathcal{\tilde{S}}_{\textit{XXZ}}[\{\rho_{n}(\lambda)\}]=\sum_{n=1}^{\infty}\int_{-\pi/2}^{\pi/2}d\lambda\rho_{n}(\lambda)\left(g_{n}^{\Psi_{0}}(\lambda)-\mu n\right)-s_{\textit{XXZ}}[\{\rho_{n}(\lambda)\}]\,.\label{eq:quenchActionXXZSpecific}
\end{equation}
To find the Bethe Ansatz root densities that describe the steady state
after the quench, the string densities $\{\rho_{n}^{*}(\lambda)\}$
minimizing $\mathcal{\tilde{S}}_{\textit{XXZ}}[\{\rho_{n}(\lambda)\}]$
have to be computed. Since (\ref{eq:quenchActionXXZSpecific}) has
the same structure as the thermal free energy functional of the \emph{XXZ}
chain in a magnetic field, the variational equations have the same
structure as those of the \emph{XXZ} thermodynamic Bethe Ansatz \cite{9780511524332}.
Introducing $\eta_{n}$ as in (\ref{eq:etadef}), the variational
equations read

\begin{equation}
\begin{aligned}\log\eta_{n}(\lambda) & =g_{n}^{\Psi_{0}}(\lambda)-\mu n+\sum_{m=1}^{\infty}\left(T_{nm}\star\log(1+\eta_{m}^{-1})\right)(\lambda)\end{aligned}
,\qquad n=1,2,...\,,\label{eq:oTbaCoupled}
\end{equation}
where the limit $\mu\rightarrow\infty$ must be taken. 

The system (\ref{eq:oTbaCoupled}) can be partially decoupled using
the properties of the functions $T_{nm}(\lambda)$ \cite{9780511524332},
as performed in Refs. \cite{Wouters2014,Brockmann2014} for quenches
starting from $|\Psi_{0}^{\text{N}}\rangle$. Since the functions
$T_{nm}(\lambda)$ are independent of the initial state, (\ref{eq:oTbaCoupled})
can be decoupled using the same method for other initial states as
well. Setting $\eta_{0}=1$, the decoupled equations take the form

\begin{equation}
\begin{aligned}\log\eta_{n}(\lambda) & =d_{n}^{\Psi_{0}}(\lambda)+[s\star\log((1+\eta_{n+1})(1+\eta_{n-1}))](\lambda)\,,\qquad n=1,2,\dots\end{aligned}
\label{eq:oTbaDecoupled}
\end{equation}
with $s(\lambda)$ defined as in (\ref{eq:sFunctionDefinition}),
and the source terms are given by 
\begin{align}
d_{n}^{\Psi_{0}}(\lambda) & =g_{n}^{\Psi_{0}}(\lambda)-[s\star(g_{n-1}^{\Psi_{0}}+g_{n+1}^{\Psi_{0}})](\lambda)\,,\qquad g_{0}^{\Psi_{0}}(\lambda)=0\,.\label{eq:oTbaDecoupledSource}
\end{align}
Explicit expressions for $d_{n}^{\Psi_{0}}(\lambda)$ can be derived
for all initial states following the steps applied for the Néel state
in \cite{Wouters2014,Brockmann2014b}:
\begin{equation}
\begin{aligned}d_{n}(\lambda) & =\xi_{1,n}^{\Psi_{0}}\log\frac{\theta_{4}^{2}(\lambda)}{\theta_{1}^{2}(\lambda)}+\xi_{2,n}^{\Psi_{0}}\log\frac{\theta_{2}^{2}(\lambda)}{\theta_{3}^{2}(\lambda)}\,,\end{aligned}
\label{eq:oTbaDecoupledSourceExplicit}
\end{equation}
where $\theta_{n}(x)$ is the $n$-th Jacobi $\theta$-function with
nome $e^{-2\eta}$, and $\xi_{1,n}^{\Psi_{0}}$ and $\xi_{2,n}^{\Psi_{0}}$
are signs depending on the particular initial state and $n$ given
in Table \ref{tab:decoupledSourceSigns}. Note that in the $q$-dimer
case the even and odd equations have the same source terms. The system
of equations (\ref{eq:oTbaDecoupledSourceExplicit}) can be solved
numerically or analytically for $\{\eta_{n}(\lambda)\}$, after which
the densities $\left\{ \rho_{n}^{*}(\lambda)\right\} $ are obtained
by solving a decoupled version of the Bethe-Takahashi equations (\ref{eq:BetheTakahashiEquations}).
The numerical solution of (\ref{eq:oTbaDecoupled}) involves the truncation
of equations in $n$ by retaining only the first $n_{\mathrm{eq}}$
equations. The truncation needs to take into account the behavior
of $\eta_{n}(\lambda)$ for $n\rightarrow\infty$ which takes the
form \cite{Brockmann2014b} 
\begin{eqnarray*}
\lim_{\begin{aligned}n & \rightarrow\infty\\
n & \mathrm{even}
\end{aligned}
}\eta_{n}(\lambda) & = & \eta_{\mathrm{even}}^{\Psi_{0}}\\
\lim_{\begin{aligned}n & \rightarrow\infty\\
n & \mathrm{even}
\end{aligned}
}\eta_{n}(\lambda) & = & \eta_{\mathrm{odd}}^{\Psi_{0}}\;,
\end{eqnarray*}
 and in the $q$-dimer case $\eta_{\mathrm{even}}^{q\mathrm{D}}=\eta_{\mathrm{odd}}^{q\mathrm{D}}$
holds. This asymptotics can be implemented in eqns. (\ref{eq:oTbaDecoupled})
by imposing $\eta_{n_{\mathrm{eq}}-2}(\lambda)=\eta_{n_{\mathrm{eq}}}(\lambda).$

The validity of the saddle-point solution $\{\rho_{n}^{*}(\lambda)\}$
is checked by
\begin{enumerate}
\item computing the expectation values of the conserved charges in the state
described by $\{\rho_{n}^{*}(\lambda)\}$ using the formula (\ref{eq:chargesIntegralFormula}).
These values should be equal to their previously known values in the
initial state which can be computed from the generating functions
(\eqref{eq:ChargeGenFunc});
\item evaluating the overlap sum rule (\ref{eq:OverlapSumRule}), which
states that the quench action (\ref{eq:quenchAction}) should be zero
at the saddle point $\{\rho_{n}^{*}(\lambda)\}$.
\end{enumerate}
For all the cases we considered, these tests were satisfied up to
the available numerical precision: the overlap sum rule gave a result
of order $10^{-8}$, while the values of the charges were reproduced
to $8-10$ digits precision.

\begin{table}
\begin{centering}
\begin{tabular}{|c||c|c|}
\hline 
$|\Psi_{0}\rangle$ & $\xi_{1,n}^{\Psi_{0}}$ & $\xi_{2,n}^{\Psi_{0}}$\tabularnewline
\hline 
\hline 
$|\Psi_{0}^{\text{N}}\rangle$ & $(-1)^{n}$ & $+1$\tabularnewline
\hline 
$|\Psi_{0}^{\text{D}}\rangle$ & $-1$ & $(-1)^{n}$\tabularnewline
\hline 
$|\Psi_{0}^{q\text{D}}\rangle$ & $-1$ & $+1$\tabularnewline
\hline 
\end{tabular}
\par\end{centering}

\protect\caption{\label{tab:decoupledSourceSigns}The signs $\xi_{1,n}^{\Psi_{0}}$
and $\xi_{2,n}^{\Psi_{0}}$ that appear in the source terms $d_{n}^{\Psi_{0}}(\lambda)$
of the decoupled form of the GTBA equations for quenches starting
from different initial states.}
\end{table}

\subsection{Exact solution to the overlap thermodynamic Bethe Ansatz equations}

In \cite{Brockmann2014b}, an exact solution was given for the Néel
oTBA equations using a functional relationship between $\eta_{1}(\lambda)$
and a function $\mathfrak{a}(\lambda)$, which is an auxiliary function
of the T-system corresponding to the Y-system of equations (\ref{eq:oTbaDecoupled}).
The relationship reads

\begin{equation}
(1+\eta_{1}(\lambda))=(1+\mathfrak{a}(\lambda+i\eta/2)\mathfrak{)}(1+\mathfrak{a}^{-1}(\lambda-i\eta/2)\mathfrak{)}.\label{eq:etaARelation}
\end{equation}
As noted in \cite{Brockmann2014b}, $\mathfrak{a}^{(\textrm{N)}}(\lambda)$
can be interpreted as the auxiliary function corresponding to the
quantum transfer matrix \cite{1993ZPhyB..91..507K,Klumper2004}. The
relation (\ref{eq:etaARelation}) is quite general: the same holds
between the corresponding auxiliary functions of the thermal T-system
and Y-system \cite{kuniba1998continued}, where the T-system is the
generalization of the thermal quantum transfer matrix \cite{9780511628832},
and the Y-system is the system of the standard thermal TBA equations
\cite{9780511524332}.

Guessing $\mathfrak{a}(\lambda)$ using the analytical structure of
(\ref{eq:oTbaDecoupled}), relation (\ref{eq:etaARelation}) gives
$\eta_{1}(\lambda)$, which is enough to solve (\ref{eq:oTbaDecoupled})
for every $\eta_{n}(\lambda)$. In the Néel case, $\mathfrak{a}(\lambda)$
was found to be \cite{Brockmann2014b}
\begin{equation}
\mathfrak{a^{(\textrm{N)}}(\lambda)}=\frac{\sin(\lambda+i\eta)\sin(2\lambda-i\eta)}{\sin(\lambda-i\eta)\sin(2\lambda+i\eta)}.\label{eq:a_function_Neel}
\end{equation}
We note that the exact solution (\ref{eq:a_function_Neel}) can be
obtained more intuitively. The expression for $\mathfrak{a(\lambda)}$
can be obtained using the boundary QTM formalism \cite{Pozsgay2013}
for the dynamical free energy density 
\[
g(s)=-\frac{1}{L}\log\langle\Psi_{0}|e^{-sH_{XXZ}(\Delta)}|\Psi_{0}\rangle\hspace{1em}(\TDL)\,.
\]
In the limit $s\rightarrow0$, $g(s)$ tends to the quench action
(\ref{eq:OverlapSumRule}), and the corresponding T-system of the
boundary QTM becomes the T-system of the quench action formalism.
Therefore the auxiliary function $\mathfrak{a}(\lambda)$ of the boundary
QTM also becomes the $\mathfrak{a}(\lambda)$ function of  \cite{Brockmann2014b}.
For $|\Psi_{0}^{\textrm{N}}\rangle$, we found that in the the $s\rightarrow0$
limit of the dynamical free energy, $\mathfrak{a}(\lambda)$ is the
function $K(u)$ of  \cite{Pozsgay2013} evaluated at $u=-i\lambda$,
which is precisely (\ref{eq:a_function_Neel}).

Following this line of thought, it is also possible to give an exact
solution of (\ref{eq:oTbaDecoupled}) for the q-dimer state $|\Psi_{0}^{q\mathrm{D}}\rangle$.
In this case the auxiliary function of the boundary QTM in the $s\rightarrow0$
limit is 
\[
\mathfrak{a}^{(q\mathrm{D})}(\lambda)=\frac{\sin(2\lambda-i\eta)}{\sin(2\lambda+i\eta)}\:,
\]
which is obtained using the function $K(u)$ of  \cite{Pozsgay2013}
corresponding to $|\Psi_{0}^{q\mathrm{D}}\rangle$. The corresponding
auxiliary function of the oTBA is, by (\ref{eq:etaARelation}):
\[
\eta_{1}^{(q\mathrm{D})}(\lambda)=\frac{\sin(2\lambda)}{\sin(2\lambda+2i\eta)}+\frac{\sin(2\lambda)}{\sin(2\lambda-2i\eta)}+\frac{\sin^{2}(2\lambda)}{\sin(2\lambda+2i\eta)\sin(2\lambda-2i\eta)}\:.
\]
This formula matches the numerical result for $\eta_{1}^{(q\mathrm{D)}}$
within the accuracy of the iterative solution. 

We note that the above argument breaks down for the dimer state $|\Psi_{0}^{\mathrm{D}}\rangle$:
the correct form of the nonlinear integral equation in the boundary
QTM formalism \cite{Pozsgay2013} is not known. It is most likely
that the problem is related to the analytic structure of the boundary
reflection factors and of the auxiliary function $\mathfrak{a}$ entering
the boundary QTM, which for the dimer state substantially differs
from the Néel and q-dimer cases.

\section{\label{sec:Computing-correlation-functions}Computing correlation
functions}

Now we turn to the evaluation of correlation functions, based on the
conjectures published in \cite{Mestyan2014,2014JSMTE..09..026P}.
These lead to the following recipe for the steady state expectation
value of short-range correlators:
\begin{enumerate}
\item Solve the equations (\ref{eq:oTbaDecoupled}), or, alternatively (\ref{eq:GGETBAEquations})
in the context of the GGE for the $\eta_{j}$.
\item With the $\eta_{j}$ thus obtained, solve the following equations
for the auxiliary functions $\rho_{n}^{(j)}(\lambda)$ and $\sigma_{n}^{(j)}(\lambda)$:
\begin{eqnarray}
\rho_{n}^{(j)}(\lambda) & = & \delta_{n,1}s^{(j)}(\lambda)+\left[s\star\left(\frac{\rho_{n-1}^{(j)}}{1+1/\eta_{n-1}}+\frac{\rho_{n+1}^{(j)}}{1+1/\eta_{n+1}}\right)\right](\lambda)\label{eq:correlationRhoEquation}\\
\sigma_{n}^{(j)}(\lambda) & = & \delta_{n,1}t^{(j)}(\lambda)+\left[t\star\left(\frac{\rho_{n-1}^{(j)}}{1+1/\eta_{n-1}}+\frac{\rho_{n+1}^{(j)}}{1+1/\eta_{n+1}}\right)\right](\lambda)+\label{eq:correlationSigmaEquation}\\
 & + & \left[s\star\left(\frac{\sigma_{n-1}^{(j)}}{1+1/\eta_{n-1}}+\frac{\sigma_{n+1}^{(j)}}{1+1/\eta_{n+1}}\right)\right](\lambda),\nonumber 
\end{eqnarray}
where
\begin{eqnarray*}
s(\lambda) & = & \frac{1}{2\pi}\left(1+2\sum_{k=1}^{\infty}\frac{\cos2k\lambda}{\cosh k\eta}\right)=\frac{d}{d\lambda}s^{(0)}(\lambda),\hspace{1em}s^{(j)}(\lambda)=\frac{d}{d\lambda}s^{(0)}(\lambda)\\
t(\lambda) & = & \mbox{\ensuremath{\frac{1}{2\pi}\sum_{k=1}^{\infty}\frac{\sinh(k\eta)}{\cosh^{2}(k\eta)}\sin(2k\lambda)}},\hspace{1em}t^{(j)}(\lambda)=\frac{d}{d\lambda}t^{(0)}(\lambda),
\end{eqnarray*}
and $\rho_{0}^{(j)}$ is defined to be $0$. The equations (\ref{eq:correlationRhoEquation})
for $\rho_{n}^{(0)}(\lambda)$ are equivalent to (\ref{eq:BetheTakahashiEquations}),
therefore $\rho_{n}^{(0)}(\lambda)$ can be identified as the total
root and hole density $\rho_{n}(\lambda)+\rho_{n}^{\mathrm{h}}(\lambda)$.
The system (\ref{eq:correlationRhoEquation},\ref{eq:correlationSigmaEquation})
is partially decoupled in the sense that the $n$th equation equation
depends only on $\rho_{n}^{(j)}(\lambda)$'s and $\sigma_{n}^{(j)}(\lambda)$'s
with three consecutive lower indices $n-1$, $n$ and $n+1$. \\
The above decoupled form of the equations appeared first in \cite{2014JSMTE..09..026P}
without derivation. In Appendix \ref{sec:appendixCorrelators}, we
show that the decoupled form is in indeed generally valid by giving
a rigorous derivation of these equations.
\item Using $\rho_{n}^{(j)}(\lambda)$ and $\sigma_{n}^{(j)}(\lambda)$,
compute the quantities
\begin{equation}
\begin{aligned}\begin{aligned}\Omega_{j,l}\end{aligned}
 & =4\pi\left[(-1)^{l}G_{j+l}+\int_{-\pi/2}^{\pi/2}d\lambda s^{(l)}(\lambda)\frac{\rho_{1}^{(j)}(\lambda)}{1+1/\eta_{1}(\lambda)}\right]\\
\Gamma_{j,l} & =4\pi\bigg[(-1)^{l}H_{j+l}-\int_{-\pi/2}^{\pi/2}d\lambda t^{(l)}(\lambda)\frac{\rho_{1}^{(j)}(\lambda)}{1+1/\eta_{1}(\lambda)}+\\
 & +\int_{-\pi/2}^{\pi/2}d\lambda s^{(l)}(\lambda)\frac{\sigma_{1}^{(j)}(\lambda)}{1+1/\eta_{1}(\lambda)}\bigg],
\end{aligned}
\label{eq:correlationIntegrals}
\end{equation}
where
\begin{eqnarray*}
G_{j} & = & -\frac{1}{\pi}\sum_{k=-\infty}^{\infty}\frac{(2ik)^{j}}{1+e^{2\eta|k|}}\\
H_{j} & =- & \frac{1}{2\pi}\sum_{k=-\infty}^{\infty}\frac{|k|(2ik)^{j-1}}{\cosh^{2}\eta k}.
\end{eqnarray*}

\item Compute the quantities
\begin{eqnarray*}
\omega_{a,b} & = & -(-1)^{(a+b)/2}\Omega_{a,b}-(-1)^{b}\frac{1}{2}\left(\frac{\partial}{\partial u}\right)^{a+b}\mathcal{K}(u)\big|_{u=0}\\
W_{a,b} & = & (-1)^{(a+b-1)/2}\Gamma_{a,b}+(-1)^{b}\frac{1}{2}\left(\frac{\partial}{\partial u}\right)^{a+b}\tilde{\mathcal{K}}(u)\big|_{u=0},
\end{eqnarray*}
with 
\begin{eqnarray*}
\mathcal{K}(u) & = & \frac{\sinh2\eta}{\sinh\left(u+\eta\right)\sinh\left(u-\eta\right)}\\
\tilde{\mathcal{K}}(u) & = & \frac{\sinh2u}{\sinh\left(u+\eta\right)\sinh\left(u-\eta\right)}.
\end{eqnarray*}

\item Substitute $\omega_{a,b}$ and $W_{a,b}$ into the QTM formulas for
short range correlations $\sigma_{j}^{z}\sigma_{j+l}^{z}$ and $\sigma_{j}^{x}\sigma_{j+l}^{x}$
that are already available in the literature \cite{2007JPhA...4010699B,2010EPJB...73..253T}.
Here we only quote the QTM formulas for the nearest neighbor and next
nearest neighbor correlations:
\end{enumerate}
\begin{equation}
\begin{split}\langle\sigma_{1}^{z}\sigma_{2}^{z}\rangle & =\coth(\eta)\omega_{0,0}+W_{1,0}\\
\langle\sigma_{1}^{x}\sigma_{2}^{x}\rangle & =-\frac{\omega_{0,0}}{2\sinh(\eta)}-\frac{\cosh(\eta)}{2}W_{1,0}\\
\langle\sigma_{1}^{z}\sigma_{3}^{z}\rangle & =2\coth(2\eta)\omega_{0,0}+W_{1,0}+\tanh(\eta)\frac{\omega_{2,0}-2\omega_{1,1}}{4}-\frac{\sinh^{2}(\eta)}{4}W_{2,1}\\
\langle\sigma_{1}^{x}\sigma_{3}^{x}\rangle & =-\frac{1}{\sinh(2\eta)}\omega_{0,0}-\frac{\cosh(2\eta)}{2}W_{1,0}-\tanh(\eta)\cosh(2\eta)\frac{\omega_{2,0}-2\omega_{1,1}}{8}+\\
 & \hspace{6cm}+\sinh^{2}(\eta)\frac{W_{2,1}}{8}.
\end{split}
\label{eq:correlationQTMFormulas}
\end{equation}

It was conjectured in \cite{Mestyan2014,2014JSMTE..09..026P} that
using the steps above, one can compute short range correlations in
an arbitrary generalized thermodynamic Bethe Ansatz state characterized
by the functions $\eta_{j}(\lambda)$. The conjecture was proved using
the Hellmann-Feynman theorem for nearest neighbor correlators, i.e.
$\sigma_{j}^{z}\sigma_{j+1}^{z}$ and $\sigma_{j}^{x}\sigma_{j+1}^{x}$
\cite{Wouters2014,Mestyan2014}, and its validity for longer range
correlators was numerically checked in the context of the standard
thermal \cite{9780511524332} thermodynamic Bethe Ansatz \cite{Mestyan2014}.

\section{Numerical results for correlations\label{sec:Numerical-results-for-correlations}}

In this section the predictions of the quench-action-based oTBA and
the GGE-based GTBA are compared to real time numerical simulations
for quenches starting from $|\Psi_{0}^{\textrm{N}}\rangle,$ $|\Psi_{0}^{\textrm{D}}\rangle$
and $|\Psi_{0}^{q\textrm{D}}\rangle$. Details of the real time simulations
are described in Appendix \ref{sec:appendixTEBD}. 

\begin{figure}
\includegraphics[width=0.45\textwidth]{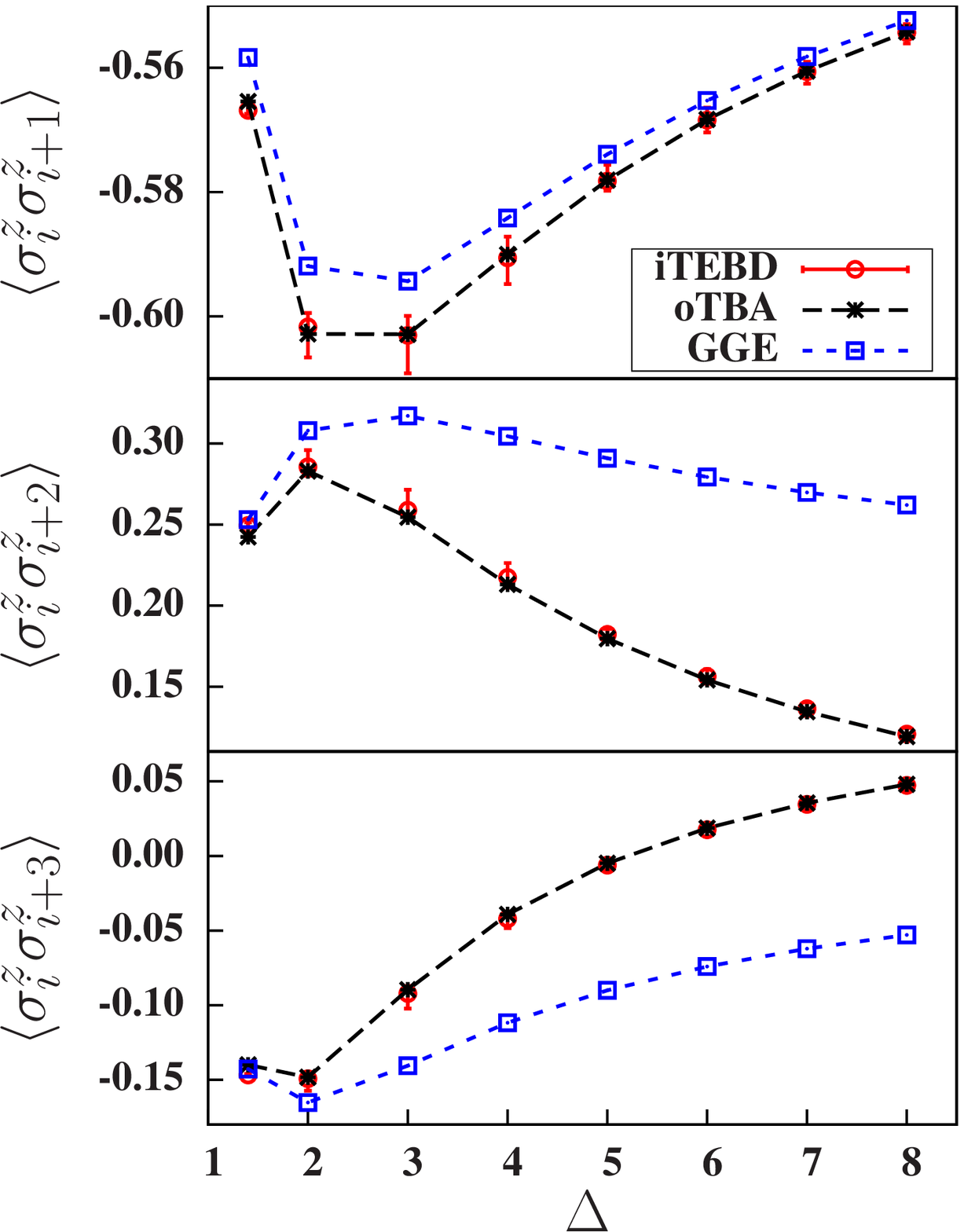}~~~~\includegraphics[width=0.45\textwidth]{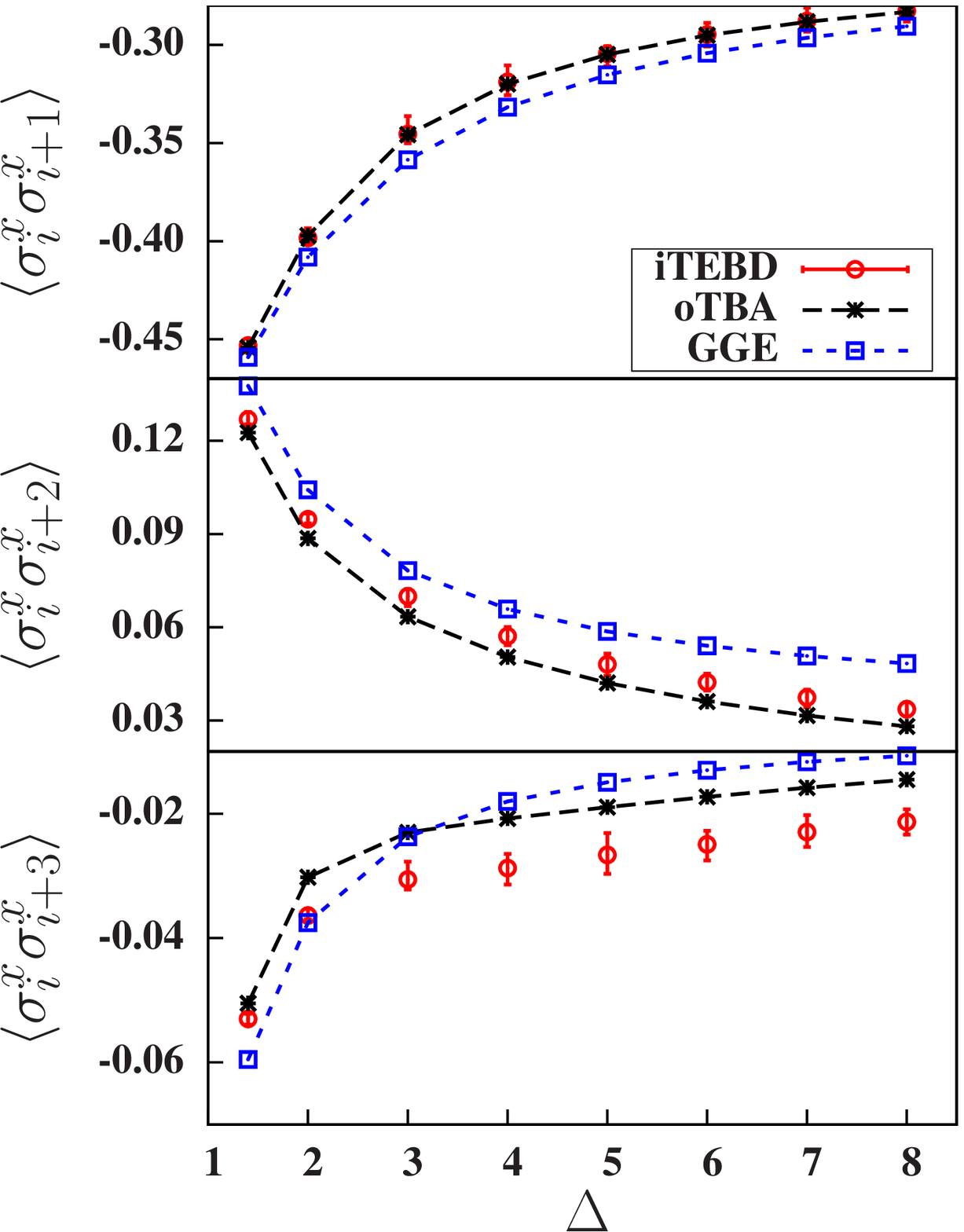}

\protect\caption{\label{fig:dimerplots}Steady state correlators for quenches starting
from the dimer initial state, as a function of the anisotropy parameter
$\Delta$.}
\end{figure}

Results for the dimer initial state are presented in Figure \ref{fig:dimerplots}.
These data are essentially the same as in our previous paper \cite{Pozsgay2014},
with the exception that we plotted the $xx$ correlations as well,
and that the GGE prediction for all the values of $\Delta$ is computed
from the full GGE using the GTBA formalism described in Subsection
\ref{sub:GTBA}. Note that while the oTBA is fully consistent with
the result numerical simulation, the GGE significantly disagrees.
The only exceptions are the correlators $\sigma_{i}^{x}\sigma_{i+2}^{x}$
and $\sigma_{i}^{x}\sigma_{i+3}^{x}$ for the dimer case, where the
real time simulation shows a temporal drift for all accessible times
(cf. Appendix \ref{sec:appendixTEBD}). For the dimer $\sigma_{i}^{x}\sigma_{i+2}^{x}$
correlator the results are still clearly consistent with the oTBA
and disagree with the GGE, but for the dimer $\sigma_{i}^{x}\sigma_{i+3}^{x}$
the numerical uncertainty introduced by the residual drift is simply
too large. 

\begin{figure}
\includegraphics[width=0.45\textwidth]{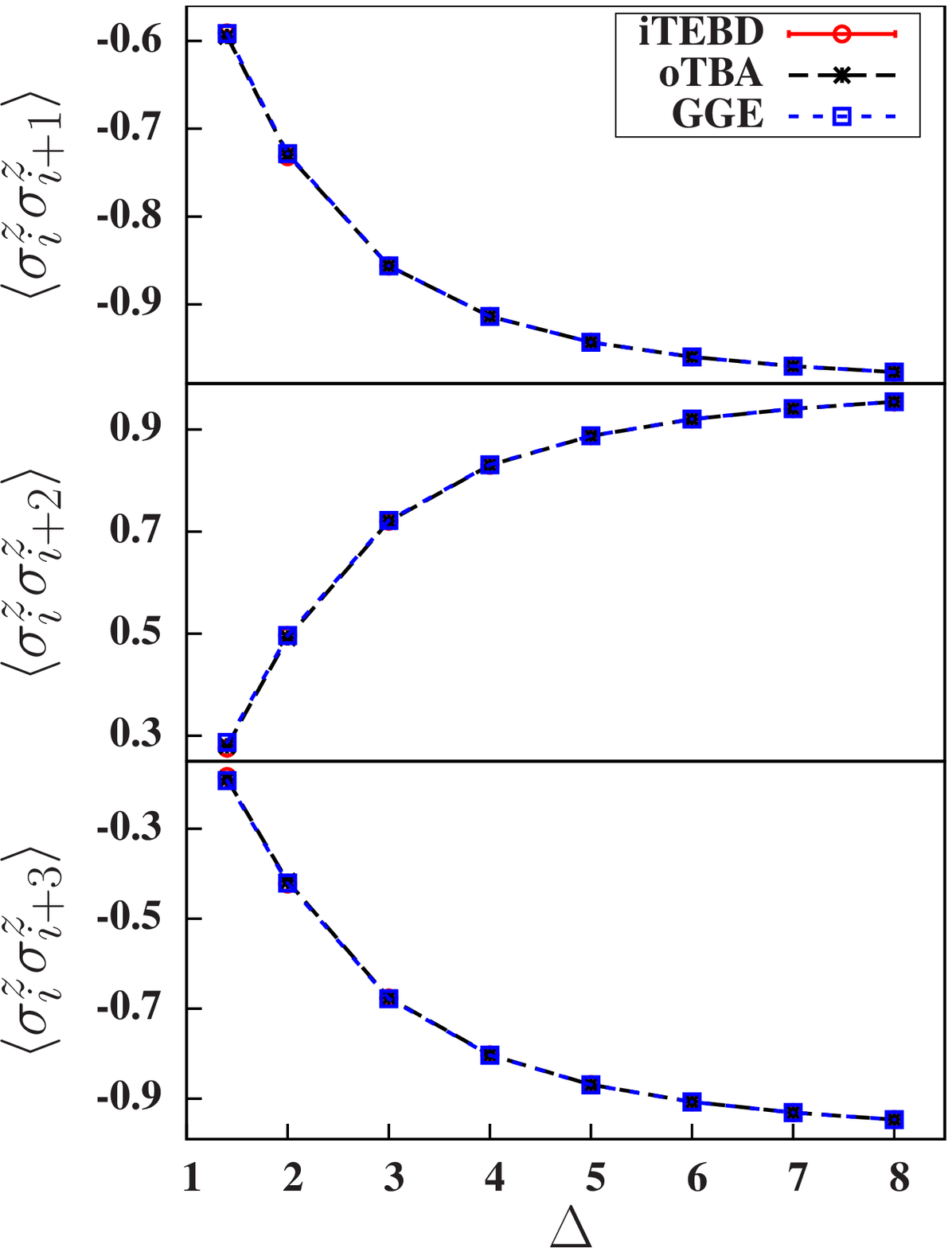}~~~~\includegraphics[width=0.45\textwidth]{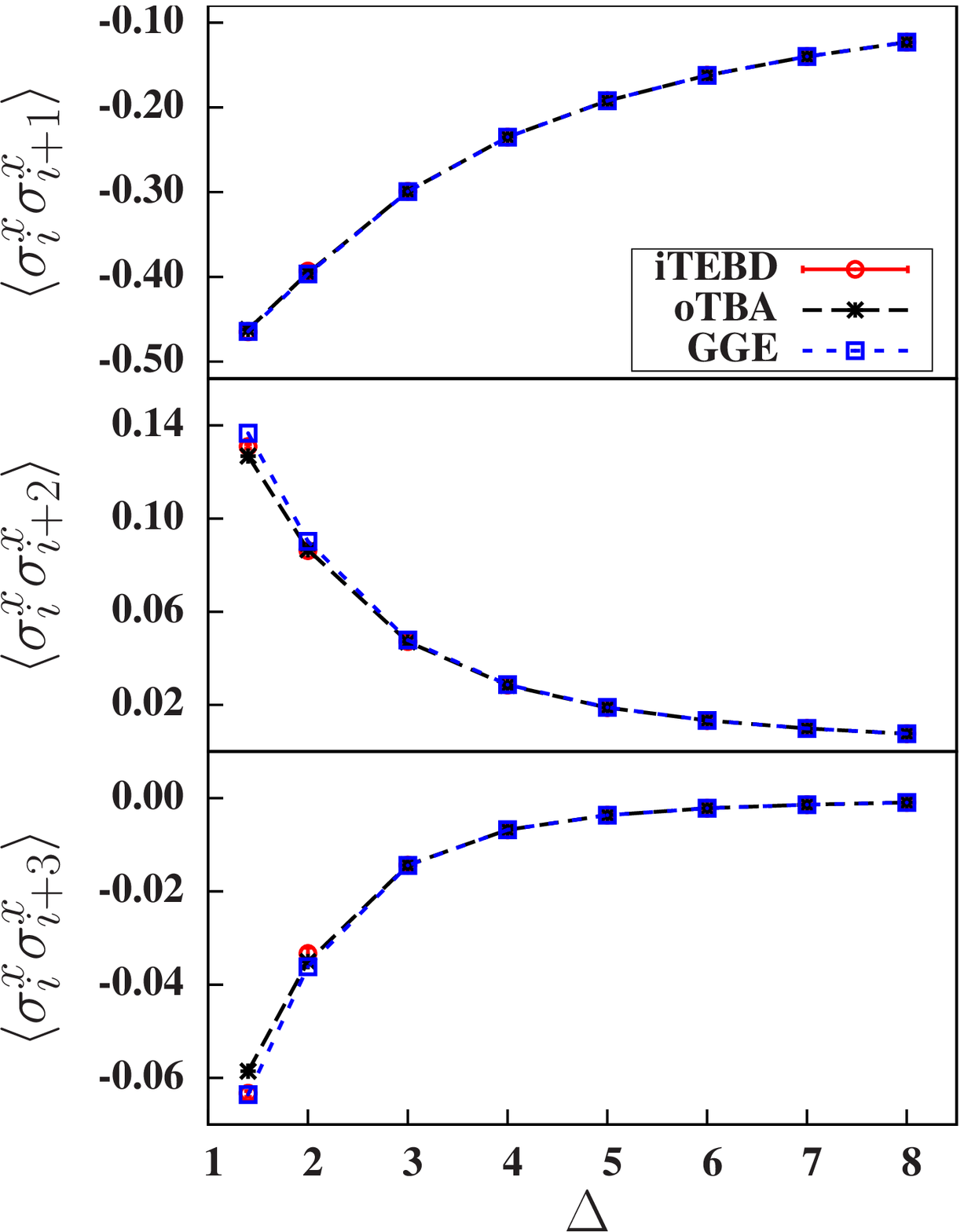}

\protect\caption{\label{fig:qdimerplots}Steady state correlators for quenches starting
from the q-dimer initial state, as a function of the anisotropy parameter
$\Delta$.}
\end{figure}

For the q-dimer case the results are shown in Figure \ref{fig:qdimerplots}.
Here the GTBA and oTBA predictions still differ, bu the difference
is too small to be resolved by the numerical simulation. We also computed
the correlators with Néel initial state, but the data are similar
to the q-dimer case, and so we omit this case, which was already discussed
in  \cite{Pozsgay2014}.

Looking at the full picture, it is clear that the conclusions of our
previous paper \cite{Pozsgay2014} still stand: the GGE clearly disagrees
with the real time evolution, while the oTBA is in agreement with
them, wherever the difference between the GGE and the oTBA is large
enough to be resolved by the numerics, and the numerically allowed
iTEBD timeframe is sufficient for the steady state to be reached.
We stress that due to the fact that the oTBA and GGE results are numerically
very close in the Néel and q-dimer cases, the dimer case plays a very
important role in deciding the issue.

\section{Discussion\label{sec:Discussion}}

The conclusion of the present work is the same as that of the previous
paper \cite{Pozsgay2014}: the GGE fails as a general description
for the steady state after quenches in the XXZ spin chain. There are,
however, several aspects which need to be examined as a possible explanation
for failure. In particular it is necessary to understand how generic
the phenomenon is: maybe the failure is due to the particular choice
of initial states, or anomalously slow relaxation? Furthermore, is
it possible that some more general or different construction for the
GGE could be adequate?

\subsection{Steady state ensemble and the role of locality \label{sub:Steady-state-ensemble-and-locality}}

We recall that the generalized Gibbs ensemble (or, for that matter,
thermalization in the non-integrable case) is not expected to describe
any observable. Indeed, as the initial state is pure, the exact state
of the system given by 
\begin{equation}
|\Psi(t)\rangle=e^{-iHt}|\Psi_{0}\rangle\label{eq:Htime_evolution}
\end{equation}
always remains pure. The relaxation towards a thermodynamic state
is always understood to work for observables defined on a particular
class of subsystems. Let us suppose that the Hilbert space of our
system can be written in the ``local'' form
\begin{equation}
\mathcal{H}=\bigotimes_{x\in S}\mathcal{V}_{x}\label{eq:local_Hilbert_space}
\end{equation}
where $x$ runs over the system $S$. For a spin chain $x$ labels
the spatial location along the chain, but this interpretation is not
strictly necessary: ``localization'' can be true in any other way,
e.g. $x$ may label particles composing a many-body system. The important
point is that the Hamiltonian governing the dynamics is supposed to
be ``local'', i.e. sum of terms consisting of product of operators
acting on a few adjacent ``local'' spaces $\mathcal{V}_{x}$, where
adjacency is some relation specifying which $x$-s are close to each
other. We also suppose that the initial state is the ground state
of some other, equally ``local'' Hamiltonian. The relevant class
of observables is then chosen to be the ``local'' ones, i.e. ones
acting on finitely many adjacent $\mathcal{V}_{x}$, and the charges
expected to play a role in the steady state statistical operator are
the ones that are ``local'' in the same way as the Hamiltonian is.
For the XXZ chain, the notion of $x$-''locality'' is just usual
spatial locality. 

The generalized Gibbs ensemble is defined as a specific thermodynamic
and temporal limit, where the order of limits matters. Let us suppose
the observable $\mathcal{O}$ is localized in the above sense, i.e.
there is a finite subsystem $O$ such that
\begin{equation}
\mathcal{O}\in\bigotimes_{x\in O}L(\mathcal{V}_{x})\label{eq:localized_observable}
\end{equation}
where $L(\mathcal{V}_{x})$ are the linear operators on $\mathcal{V}_{x}$;
let us call the smallest such subsystem the support of $\mathcal{O}$.
Then the statistical operator $\rho_{\mbox{SS}}$ describes the steady
state if it is true that
\begin{equation}
\Tr\rho_{\mbox{SS}}\mathcal{O}=\lim_{t\rightarrow\infty}\TDL\,\langle\Psi(t)|\mathcal{O}|\Psi(t)\rangle\label{eq:steady_state_definition}
\end{equation}
where $\TDL$ is the thermodynamic limit in which the size of the
system $S$ goes to infinity. In fact it is possible to allow slightly
more generality and allow the support of $\mathcal{O}$ to grow in
the thermodynamic limit provided the system size becomes infinitely
larger than the support of $\mathcal{O}$.

The GGE hypothesis can be stated as 
\begin{eqnarray}
\Tr\rho_{\mbox{SS}}\mathcal{O} & = & \lim_{N\rightarrow\infty}\Tr\rho_{N}\mathcal{O}\nonumber \\
\rho_{N} & = & \frac{1}{Z_{N}}\exp\!\left(\sum\limits _{k=1}^{N}\beta_{k}Q_{k}\right)\qquad Z_{N}=\Tr\exp\!\left(\sum\limits _{k=1}^{N}\beta_{k}Q_{k}\right)\label{eq:GGE_from_TGGE}
\end{eqnarray}
Note that the GGE is defined above as the limit of the truncated GGE
introduced in \cite{2013PhRvB..87x5107F,Pozsgay2013b}. This is motivated
by the fact that the infinite sum in the exponent needs some proper
definition, especially since including terms up to $N=\infty$ is
not really local. It is further justified by the observation that
the truncated GGE approximates the full one well; in fact, as described
in Subsection \ref{sub:GTBA}, charges that are much ``larger''
than the support of $\mathcal{O}$ do not have much influence on the
GGE prediction for the expectation value of $\mathcal{O}$.

The above description generally captures the way the GGE was expected
to work \emph{for systems with local dynamics} such as the XXZ spin
chain \cite{Pozsgay2013b,Fagotti2014}. What we have shown is that
\emph{the GGE defined in the above way definitely fails to describe
the steady state of the system}.

\subsection{Failure of the Generalized Eigenstate Thermalization Hypothesis}

Soon after the results of the papers \cite{Wouters2014,Pozsgay2014}
were made public, work started to explore the mechanism responsible
for the failure of the GGE. In \cite{Goldstein2014} it was pointed
out that the expectation values of the initial charges did not specify
the string densities $\rho_{n}$ uniquely. The reason is the existence
of multiple types of strings, which means that the elementary magnetic
excitations of the XXZ chain, described by 1-strings, have bound states
corresponding to longer strings. While the GGE corresponds to maximizing
the entropy among configurations allowed by the particular values
of the charges specified by the initial state, this does not lead
to the same solution that follows from the oTBA, and generally gives
different correlations. Indeed this particular point is very interesting,
and was investigated further in \cite{2014JSMTE..09..026P}. The fact
that expectation values of local operators in the thermodynamic limit
is not specified by the values of local charges means the failure
of the Generalized Eigenstate Thermalization Hypothesis (GETH). 

Let us recall how the usual Eigenstate Thermalization Hypothesis works
for generic (i.e. non-integrable systems) \cite{1991PhRvA..43.2046D,1994PhRvE..50..888S,2008Natur.452..854R}.
As discussed in Subsection \ref{sub:The-diagonal-ensemble}, in the
large time limit (neglecting degeneracies) one obtains the prediction
of the diagonal ensemble (\ref{eq:diagonalEnsemble}), where each
state is weighted by the squared norm of its overlap with the initial
state. If the system reaches a thermal equilibrium, then the expectation
values of relevant operators should be close to the canonical prediction
\begin{equation}
\left\langle \mathcal{O}\right\rangle _{T}=\frac{\sum_{n}e^{-E_{n}/T}\langle n|\mathcal{O}|n\rangle}{\sum_{n}e^{-E_{n}/T}}\label{micro}
\end{equation}
with a temperature $T$ that is fixed by the requirement 
\begin{equation}
\left\langle H\right\rangle _{T}=\langle\Psi_{0}|H|\Psi_{0}\rangle\;.
\end{equation}
The diagonal ensemble (\eqref{eq:diagonalEnsemble}) and the thermal
averages \eqref{micro} are expected to become equal in the $\TDL$.
The overlap coefficients entering (\eqref{eq:diagonalEnsemble}) are
typically random and different from the Boltzmann weights. However,
in a large volume $L$ the only states with non-negligible overlap
are the ones that have the same energy density as the initial state
\cite{2008Natur.452..854R}: 
\begin{equation}
\frac{E_{n}}{L}\approx\frac{\langle\Psi_{0}|H|\Psi_{0}\rangle}{L}\:,\label{e1}
\end{equation}
and the width of the distribution of the energy density goes to zero
in the $\TDL$: 
\begin{equation}
\Delta\left(\frac{E}{L}\right)=\frac{1}{L}\sqrt{\langle\Psi_{0}|H^{2}|\Psi_{0}\rangle-(\langle\Psi_{0}|H|\Psi_{0}\rangle)^{2}}\sim\frac{1}{\sqrt{L}}\:.\label{width}
\end{equation}
for local Hamiltonians and initial states $\ket{\Psi_{0}}$ satisfying
the cluster decomposition principle.

The \emph{Eigenstate Thermalization Hypothesis} (ETH) \cite{1991PhRvA..43.2046D,1994PhRvE..50..888S,2008Natur.452..854R}
states that eigenstates on a given energy shell have almost the same
expectation values of physical observables 
\begin{equation}
\sum_{n}|c_{n}|^{2}\langle n|\mathcal{O}|n\rangle\approx\left(\sum_{n}|c_{n}|^{2}\right)\langle n_{1}|\mathcal{O}|n_{1}\rangle=\langle n_{1}|\mathcal{O}|n_{1}\rangle\:,\label{ETH}
\end{equation}
where $n_{1}$ is a reference state satisfying condition \eqref{e1}
and $c_{1}\ne0$. Recent results \cite{2014arXiv1408.0535K} indicate
that (\eqref{ETH}) holds in the strict sense that even the largest
deviations from the ETH go to zero in the $\TDL$, albeit some local
operators may have anomalously slow relaxation rates \cite{2014arXiv1410.4186K}.
From the ETH it follows that the local observables indeed thermalize:
\begin{equation}
\lim_{t\to\infty}\langle\Psi(t)|\mathcal{O}|\Psi(t)\rangle\approx\left\langle \mathcal{O}\right\rangle _{T}\:,\label{thermalization}
\end{equation}
where the equality is expected to become exact in the $\TDL$.

For integrable systems the steady state was expected to be described
by the generalized Gibbs ensemble discussed in Subsection \eqref{sub:GTBA}.
A possible mechanism for the relaxation to GGE is provided by the
\emph{Generalized Eigenstate Thermalization Hypothesis} (GETH) \cite{2011PhRvL.106n0405C},
which states that if all local conserved charges of two different
eigenstates are close to each other, then the mean values of local
operators are also close. Put differently the values of the conserved
charges uniquely determine the correlations in the state; more precisely
this is expected to become exact in the $\TDL$. The GETH was checked
for a lattice model of hard-core bosons in \cite{2011PhRvL.106n0405C}.

However, in \cite{2014JSMTE..09..026P} it was shown that\emph{ the
GETH fails in the XXZ spin chain, and it was argued that this is the
general case for models with bound states}. In \cite{Pozsgay2014c}
it was shown that, on the other hand, the GGE holds in a system of
strongly interacting bosons with no bound states. As noted in that
work, even though the GGE hypothesis was confirmed, it was not eventually
used to describe the steady state. The crucial ingredient was the
assumption that the Diagonal Ensemble is valid, plus the fact that
there is a one-to-one correspondence between the charges and the root
densities, from which the validity of the GGE follows automatically.
The steady state correlators were evaluated from the Bethe root density
that was obtained directly from the charges using the analogue of
the relation (\ref{eq:generatingFunctionRho1h}) for the q-boson model,
with the GTBA equations described in Subsection \eqref{sub:GTBA}
playing no role whatsoever. It seems that this behaviour might be
a generic feature of interacting integrable models: \emph{the GETH
only holds when there is a one-to-one correspondence between the charges
and the root densities}. In such a case, the GGE density matrix gives
the correct steady state correlators for operators whose expectation
values depend only on the root densities, since the solution of the
GTBA coincides with the unique allowed root densities for the given
initial value of the conserved charges. 

On the other hand, if the GETH does not hold, then the GTBA analysis
of the GGE density matrix is expected to give wrong predictions for
a generic initial state. This shows that\emph{ the failure of the
GGE observed in \cite{Wouters2014,Pozsgay2014} and the present work
is not related to the selection of the initial states}. In addition,
the fact that the oTBA predictions agree with the numerical simulations
shows that relaxation to the diagonal ensemble has been achieved in
the time frame for which the iTEBD simulations are valid (with the
exception of some $\sigma^{x}\sigma^{x}$ correlators, cf. Appendix
\eqref{sec:appendixTEBD}), and therefore \emph{the disagreement with
GGE cannot be due to some anomalously slow relaxation}. 

We remark that our results (see Fig. \eqref{zzThermalization}) are
consistent with the observation made in \cite{Fagotti2014} that translational
invariance is not restored on the observable time scale for quenches
starting from the non-translationally invariant dimer state, which
affects correlators $\sigma_{i}\sigma_{i+l}$ with $l$ odd. It is
possible that translational invariance in these cases is restored
on some anomalously long time scale, as argued in \cite{Fagotti2014a}.

Another interesting observation made in \cite{2014JSMTE..09..026P}
that the quench action (\eqref{eq:quenchAction}) evaluated at the
GGE saddle point solution is infinite, and the spectral weight of
the GGE solution (and in fact of all states with thermal asymptotics)
decays faster than exponential as a function of the volume. This means
that the sum rule (\eqref{eq:OverlapSumRule}) is violated, and \emph{the
GGE solution is very far from the states selected by the quench action,
which actually determine the dynamics of the system}.

\subsection{Completeness of the charges and possible extensions of the GGE}

The discussion of the GETH shows that \emph{one way to construct the
correct ensemble is to find charges which fix all Bethe root densities
to their correct values}, thus making the GGE complete and correct.
In this work the GGE was constructed using the local charges obtained
as derivatives of the transfer matrix. 

However, new quasi-local operators have been found recently for the
regime $\Delta<1$ \cite{2011PhRvL.106u7206P,2014NuPhB.886.1177P,2014JSMTE..09..037P}
and it is interesting open question whether the addition of these
new operators to the GGE is enough to fix all root densities to their
correct values, thus obtaining an extended GGE which can describe
the steady state after the quantum quench. Unfortunately the construction
of \cite{2011PhRvL.106u7206P,2014NuPhB.886.1177P,2014JSMTE..09..037P}
does not produce quasi-local operators for $\Delta>1$, and it remains
to be seen whether new charges can be produced by other means in this
regime.

It is also important to keep in mind that consistency of a thermodynamic
description requires extensivity of the charges involved in the ensemble.
Locality is one way of ensuring extensivity, but more general (e.g.
quasi-local) charges could be suitable for inclusion into a thermodynamic
ensemble. In free systems, for example, the GGE is usually formulated
in terms of the mode occupation numbers, which are non-local quantities. 

Recently the extension of the GGE by quasi-local charges was considered
for quantum field theories; however, the charges discussed in \cite{2014arXiv1411.5352E}
are already included in our set of local charges at the level of the
discrete chain. At present it is an open question whether there exists
a set of charges, possibly including quasi-local ones that would be
complete in the sense of fixing the root densities. In the light of
recent interest in experimental observation of the GGE \cite{2014arXiv1411.7185L}\emph{
it would also be preferable if the extension preserved the truncatability
of the GGE} (cf. \cite{2013JSMTE..07..003P}, see also Subsections
\ref{sub:GTBA} and \ref{sub:Steady-state-ensemble-and-locality}),
namely that truncation to a small subset of charges would give a reasonable
approximation for the exact result, so that fitting the ensemble to
measurements would only require a small number of parameters.

\paragraph*{Acknowledgments}

We would like to thank M. Kormos and G. Zaránd for numerous discussions,
their valuable feedback, and F. Pollmann and P. Moca for the help
they gave us to set up the iTEBD calculations. We are grateful to
J.-S. Caux, J. De Nardis, and B. Wouters for useful discussions on
their work, and for sharing some of their numerical data with us.
M.A.W. acknowledges financial support from Hungarian Grants No. K105149
and CNK80991.

\bibliographystyle{bib/utphys}
\bibliography{bib/XXZ_quench_long}
\newpage{}

\begin{appendices}

\section{Mean values of correlators: decoupling the equations\label{sec:appendixCorrelators}}

In this appendix we show that the partially decoupled equations (\ref{eq:correlationRhoEquation}-\ref{eq:correlationSigmaEquation}),
which first appeared in \cite{2014JSMTE..09..026P}, are indeed equivalent
to the original coupled equations of \cite{Mestyan2014}, which read

\begin{eqnarray}
\rho_{n}^{(j)}(\lambda) & = & a_{n}^{(j)}(\lambda)-\sum_{m=1}^{\infty}T_{nm}\star\frac{\rho_{m}^{(j)}}{1+\eta_{m}}\label{eq:correlationRhoEquationCoupled}\\
\sigma_{n}^{(j)}(\lambda) & = & -b_{n}^{(j)}(\lambda)+\sum_{m=1}^{\infty}U_{nm}\star\frac{\rho_{m,\mbox{t}}^{(j)}}{1+\eta_{m}}-\sum_{m=1}^{\infty}T_{nm}\star\frac{\sigma_{m}^{(j)}}{1+\eta_{m}}\,,\label{eq:correlationSigmaEquationCoupled}
\end{eqnarray}
with the convolution kernels being
\[
T_{nm}=(1-\delta_{n,m})a_{|n-m|}+a_{n+m}+2\sum_{j=1}^{\min(n,m)-1}a_{|n-m|+2j}
\]
\[
U_{nm}=(1-\delta_{n,m})b_{|n-m|}+b{}_{n+m}+2\sum_{j=1}^{\min(n,m)-1}b_{|n-m|+2j}\,,
\]
and the source terms being
\begin{eqnarray}
a_{n}^{(j)}(\lambda) & = & \left(\frac{d}{d\lambda}\right)^{j}a_{n}(\lambda)\label{eq:a_njDefinition}\\
b_{n}(\lambda) & = & \frac{1}{2\pi}i\frac{d}{d\eta}\log\left(\frac{\sin(\lambda+in\eta/2)}{\sin(\lambda-in\eta/2)}\right)=-\frac{n}{2\pi}\frac{\sin(2\lambda)}{\cosh n\eta-\cos(2\lambda)}\frac{\sigma_{n+1}^{(j)}}{1+1/\eta_{n+1}}\nonumber \\
b_{n}^{(j)}(\lambda) & = & \left(\frac{d}{d\lambda}\right)^{j}b_{n}(\lambda)\,,\label{eq:b_njDefinition}
\end{eqnarray}
where $a_{n}(\lambda)$ is defined by (\ref{eq:a_nDefinition}). 

In order to prove the equivalence of (\ref{eq:correlationRhoEquationCoupled}-\ref{eq:b_njDefinition})
to (\ref{eq:correlationRhoEquation}-\ref{eq:correlationSigmaEquation}),
it is convenient to introduce the following notations for the Fourier
transform

\[
\tilde{f}(k)=(\mathcal{F}f)(k)=\int_{-\pi/2}^{\pi/2}d\lambda f(\lambda)e^{2ik\lambda}\qquad k\in\mathbb{Z}\,.
\]
For $k\in\mathbb{Z},$ $\tilde{f}(k)$ is the $k$-th Fourier coefficient
of $f$:

\[
f(\lambda)=(\mathcal{F}^{-1}\tilde{f})(\lambda)=\frac{1}{\pi}\sum_{k\in\mathbb{Z}}\tilde{f}(k)e^{-2ik\lambda}.
\]
The Fourier transforms of the functions (\ref{eq:a_njDefinition})
are given by

\[
\tilde{a}_{n}^{(j)}(k)=(-2ik)^{j}e^{-n\eta|k|},
\]
while the Fourier transforms of (\ref{eq:b_njDefinition}) have the
form

\begin{equation}
\tilde{b}_{n}^{(j)}(k)=\frac{-in\sgn(k)(-2ik)^{j}}{2}e^{-n\eta|k|}=\frac{i}{2\eta}\frac{d}{dk}\tilde{a}_{n}^{(j)}(k)-\frac{j}{2\eta}\tilde{a}_{n}^{(j-1)}(k)\,,\label{eq:correlationSourceRelation}
\end{equation}
where $\sgn(0)=0$ and $\sgn(x)=x/|x|$ for $x\neq0$. Although (\ref{eq:a_njDefinition})
and (\ref{eq:b_njDefinition}) define $a_{n}(\lambda)$ and \textbf{$b_{n}(\lambda)$}
for $n>1$ only, it is useful to define $\tilde{a}_{0}^{(j)}(k)$
and $\tilde{b}_{0}^{(j)}(k)$ by extending the equations to $n=0$.
We also define $\tilde{a}_{n}(k)=\tilde{a}_{n}^{(0)}(k)$ and $\tilde{b}_{n}(k)=\tilde{b}_{n}^{(0)}(k)$.

Using the above notations, the Fourier transforms of the coupled equations
(\ref{eq:correlationRhoEquationCoupled}-\ref{eq:correlationSigmaEquationCoupled})
are the following
\begin{eqnarray}
\rho_{n}^{(j)}(k) & = & \tilde{a}_{n}^{(j)}(k)-\sum_{m=1}^{\infty}\tilde{T}_{nm}(k)\left(\mbox{\ensuremath{\mathcal{F}}}\frac{\rho_{m}^{(j)}}{1+\eta_{m}}\right)(k)\label{eq:correlationRhoEquationFourier}\\
\sigma_{n}^{(j)}(k) & = & -\tilde{b_{n}}^{(j)}(k)+\sum_{m=1}^{\infty}\tilde{U}_{nm}(k)\left(\mathcal{F}\frac{\rho_{m}^{(j)}}{1+\eta_{m}}\right)(k)-\sum_{m=1}^{\infty}\tilde{T}_{nm}(k)\left(\mathcal{F}\frac{\sigma_{m}^{(j)}}{1+\eta_{m}}\right)(k)\,.\label{eq:correlationSigmaEquationFourier}
\end{eqnarray}
The decoupled form of (\ref{eq:correlationRhoEquationFourier}) is
obtained by subtracting the sum of $n-1$th and $n+1$th equations
multiplied by $\tilde{a}_{1}(k)$, from the $n$th equation multiplied
by $\tilde{a}_{0}(k)+\tilde{a}_{2}(k)$:

\begin{equation}
(\tilde{a}_{0}(k)+\tilde{a}_{2}(k))\rho_{n}^{(j)}(k)=\tilde{a}_{1}(k)\delta_{n,1}+\tilde{a}_{1}(k)\left[\left(\mathcal{F}\frac{\rho_{n-1}^{(j)}}{1+1/\eta_{n-1}}\right)(k)+\left(\mathcal{F}\frac{\rho_{n+1}^{(j)}}{1+1/\eta_{n+1}}\right)(k)\right]\,.\label{eq:correlationRhoDecoupledFourier}
\end{equation}
Introducing 
\[
\tilde{s}(k)=\frac{1}{\tilde{a}_{1}(k)+\tilde{a}_{-1}(k)}
\]
we obtain

\[
\tilde{\rho}_{n}^{(j)}(k)=\tilde{s}(k)\delta_{n,1}+\tilde{s}(k)\left[\left(\mathcal{F}\frac{\rho_{n-1}^{(j)}}{1+1/\eta_{n-1}}\right)(k)+\left(\mathcal{F}\frac{\rho_{n+1}^{(j)}}{1+1/\eta_{n+1}}\right)(k)\right]\,,
\]
which is the Fourier transform of the partially decoupled equation
(\ref{eq:correlationRhoEquation}).

Decoupling (\ref{eq:correlationSigmaEquationFourier}) proceeds via
a relation between the Fourier transforms of the convolution kernels
$\tilde{T}_{nm}(k)$ and $\tilde{U}_{nm}(k)$ similar to (\ref{eq:correlationSourceRelation}):

\begin{equation}
\tilde{U}_{nm}(k)=\frac{i}{2\eta}\frac{d}{dk}\tilde{T}_{nm}(k)\,.\label{eq:correlationKernelRelation}
\end{equation}
Acting with $(i/2\eta)(d/dk)$ on (\ref{eq:correlationRhoEquationFourier}),
we obtain
\begin{equation}
\begin{aligned}0= & \frac{i}{2\eta}\frac{d}{d(k)}\tilde{a}_{n}(k)-\sum_{m=1}^{\infty}\frac{i}{2\eta}\left(\frac{d}{dk}\tilde{T}_{nm}(k)\right)\left(\mathcal{F}\frac{\rho_{m}^{(j)}}{1+\eta_{m}}\right)(k)-\\
- & \frac{i}{2\eta}\left[\frac{d}{dk}\rho_{n}^{(j)}(k)+\sum_{m=1}^{\infty}\tilde{T}_{nm}(k)\frac{d}{dk}\left(\mathcal{F}\frac{\rho_{m}^{(j)}}{1+\eta_{m}}\right)(k)\right]\,.
\end{aligned}
\label{eq:correlationIntermediateFormula-2}
\end{equation}
Using the relations (\ref{eq:correlationSourceRelation}) and (\ref{eq:correlationKernelRelation})
on (\ref{eq:correlationSigmaEquationFourier}), we get

\begin{equation}
\begin{aligned}\sigma_{n}^{(j)}(k) & =-\frac{i}{2\eta}\frac{d}{d(k)}\tilde{a}_{n}(k)+\sum_{m=1}^{\infty}\frac{i}{2\eta}\left(\frac{d}{dk}\tilde{T}_{nm}(k)\right)\left(\mathcal{F}\frac{\rho_{m}^{(j)}}{1+\eta_{m}}\right)(k)+\\
 & +\frac{j}{2\eta}\tilde{a}_{n}^{(j-1)}(k)-\sum_{m=1}^{\infty}\tilde{T}_{nm}(k)\left(\mathcal{F}\frac{\sigma_{m}^{(j)}}{1+\eta_{m}}\right)(k)\,.
\end{aligned}
\label{eq:correlationIntermediateFormula-3}
\end{equation}
Substituting the first two terms of (\ref{eq:correlationIntermediateFormula-3})
with the third term of (\ref{eq:correlationIntermediateFormula-2})
yields

\begin{equation}
\begin{aligned}\sigma_{n}^{(j)}(k) & =-\frac{i}{2\eta}\left[\frac{d}{dk}\rho_{n}^{(j)}(k)+\sum_{m=1}^{\infty}\tilde{T}_{nm}(k)\frac{d}{dk}\left(\mathcal{F}\frac{\rho_{m}^{(j)}}{1+\eta_{m}}\right)(k)\right]+\\
 & +\frac{j}{2\eta}\tilde{a}_{n}^{(j-1)}(k)-\sum_{m=1}^{\infty}\tilde{T}_{nm}(k)\left(\mathcal{F}\frac{\sigma_{m}^{(j)}}{1+\eta_{m}}\right)(k).
\end{aligned}
\label{eq:correlationIntermediateFormula-4}
\end{equation}
Now the system (\ref{eq:correlationIntermediateFormula-4}) is easily
decoupled, similarly to (\ref{eq:correlationRhoEquationFourier}).
Subtracting the sum of $n-1$th and $n+1$th equations multiplied
by $\tilde{a}_{1}(k)$, from the $n$th equation multiplied by $\tilde{a}_{0}(k)+\tilde{a}_{2}(k)$
leads to

\begin{equation}
\begin{aligned}(\tilde{a}_{0}(k)+\tilde{a}_{2}(k))\tilde{\sigma}_{n}^{(j)}(k) & =-\frac{i}{2\eta}\Bigg\{(\tilde{a}_{0}(k)+\tilde{a}_{2}(k))\frac{d}{dk}\tilde{\rho}_{n}^{(j)}(k)\,-\\
 & -\tilde{a}_{1}(k)\left[\frac{d}{dk}\tilde{\rho}_{n-1}^{(j)}(k)+\frac{d}{dk}\tilde{\rho}_{n+1}^{(j)}(k)\right]+\\
 & +\tilde{a}_{1}(k)\left[\frac{d}{dk}\left(\mathcal{F}\frac{\rho_{n-1}^{(j)}}{1+\eta_{n-1}}\right)(k)+\frac{d}{dk}\left(\mathcal{F}\frac{\rho_{n+1}^{(j)}}{1+\eta_{n+1}}\right)(k)\right]\Bigg\}\,+\\
 & +\frac{j}{2\eta}\tilde{a}_{1}^{(j-1)}(k)\delta_{n,1}\,+\\
 & +\tilde{a}_{1}(k)\left[\left(\mathcal{F}\frac{\sigma_{n-1}^{(j)}}{1+1/\eta_{n-1}}\right)(k)+\left(\mathcal{F}\frac{\sigma_{n+1}^{(j)}}{1+1/\eta_{n+1}}\right)(k)\right]\,.
\end{aligned}
\label{eq:correlationIntermediateFormula-1}
\end{equation}
This system is already partially decoupled like (\ref{eq:correlationRhoDecoupledFourier})
but it can be simplified even further. Rearranging the terms in the
curly brackets leads to
\begin{equation}
\begin{aligned}(\tilde{a}_{0}+\tilde{a}_{2})\sigma_{n}(k) & =-\frac{i}{2\eta}\Bigg\{(\tilde{a}_{0}(k)+\tilde{a}_{2}(k))\frac{d}{dk}\rho_{n}^{(j)}(k)-\\
 & -\tilde{a}_{1}(k)\left[\frac{d}{dk}\left(\mathcal{F}\frac{\rho_{n-1}^{(j)}}{1+1/\eta_{n-1}}\right)(k)+\,\frac{d}{dk}\left(\mathcal{F}\frac{\rho_{n+1}^{(j)}}{1+1/\eta_{n+1}}\right)(k)\right]\Bigg\}\,+\\
 & +\frac{j}{2\eta}\tilde{a}_{1}^{(j-1)}(k)\delta_{n,1}\,+\\
 & +\tilde{a}_{1}(k)\left[\left(\mathcal{F}\frac{\sigma_{n-1}^{(j)}}{1+1/\eta_{n-1}}\right)(k)+\left(\mathcal{F}\frac{\sigma_{n+1}^{(j)}}{1+1/\eta_{n+1}}\right)(k)\right]\,.
\end{aligned}
\label{eq:correlationIntermediateFormula-5}
\end{equation}
Using the Leibniz rule $f(dg/dk)=d(fg)/dk-(df/dk)g$ and formula (\ref{eq:correlationRhoDecoupledFourier}),
the expression between the curly brackets can be written as

\[
\begin{aligned} & \left[\frac{d}{d(k)}\tilde{a}_{1}^{(j)}(k)-\tilde{a}_{1}^{(j)}(k)\frac{\frac{d}{dk}\tilde{a}_{0}(k)+\frac{d}{dk}\tilde{a}_{2}(k)}{\tilde{a}_{0}(k)+\tilde{a}_{2}(k)}\right]\delta_{n,1}+\\
+ & \left[\frac{d}{d(k)}\tilde{a}_{1}(k)-\tilde{a}_{1}(k)\frac{\frac{d}{dk}\tilde{a}_{0}(k)+\frac{d}{dk}\tilde{a}_{2}(k)}{\tilde{a}_{0}(k)+\tilde{a}_{2}(k)}\right]\left[\left(\mathcal{F}\frac{\rho_{n-1}^{(j)}}{1+1/\eta_{n-1}}\right)(k)+\left(\mathcal{F}\frac{\rho_{n+1}^{(j)}}{1+1/\eta_{n+1}}\right)(k)\right]\,.
\end{aligned}
\]
Introducing the notation

\[
\begin{aligned}\tilde{t}^{(j)}(k) & =-\frac{\tilde{b}_{1}^{(j)}(k)-\tilde{a}_{1}^{(j)}(k)\frac{\tilde{b}_{0}(k)+\tilde{b}_{2}(k)}{\tilde{a}_{0}(k)+\tilde{a}_{2}(k)}}{\tilde{a}_{0}(k)+\tilde{a}_{2}(k)}=\frac{i(-2ik)^{j}}{2}\frac{e^{\eta|k|}-e^{-\eta|k|}}{(e^{\eta|k|}+e^{-\eta|k|})^{2}}\end{aligned}
\]
the system (\ref{eq:correlationIntermediateFormula-5}) transforms
to
\begin{eqnarray*}
\tilde{\sigma}_{n}^{(j)}(k) & = & \delta_{n,1}\tilde{t}^{(j)}(k)+\tilde{t}(k)\left[\mathcal{F}\left(\frac{\rho_{n-1}^{(j)}}{1+1/\eta_{n-1}}\right)(k)+\left(\mathcal{F}\frac{\rho_{n+1}^{(j)}}{1+1/\eta_{n+1}}\right)(k)\right]+\\
 & + & \tilde{s}(k)\left[\left(\mathcal{F}\frac{\sigma_{n-1}^{(j)}}{1+1/\eta_{n-1}}\right)(k)+\left(\mathcal{F}\frac{\sigma_{n+1}^{(j)}}{1+1/\eta_{n+1}}\right)(k)\right],
\end{eqnarray*}
 which is precisely the Fourier transform of the system (\ref{eq:correlationSigmaEquation}).

Finally, formulas (\ref{eq:correlationIntegrals}) can be obtained
directly from their coupled counterparts \cite{Mestyan2014}
\[
\begin{aligned}\Omega_{j,l} & =4\pi\sum_{n=1}^{\infty}\int_{-\pi/2}^{\pi/2}d\lambda a_{n}^{(l)}(\lambda)\frac{\rho_{n}^{(j)}(\lambda)}{1+\eta_{n}(\lambda)}\\
\Gamma_{j,l} & =-4\pi\sum_{n=1}^{\infty}\int_{-\pi/2}^{\pi/2}d\lambda\left(b_{n}^{(l)}(\lambda)\frac{\rho^{(j)}(\lambda)}{1+\eta_{n}(\lambda)}+a_{n}^{(l)}(\lambda)\frac{\sigma^{(j)}(\lambda)}{1+\eta_{n}(\lambda)}\right)
\end{aligned}
\]
by using the partially decoupled equations (\ref{eq:correlationRhoEquation}-\ref{eq:correlationSigmaEquation}).

\section{Details of iTEBD simulations\label{sec:appendixTEBD}}

\begin{figure}
\centering{}\includegraphics[width=0.48\textwidth]{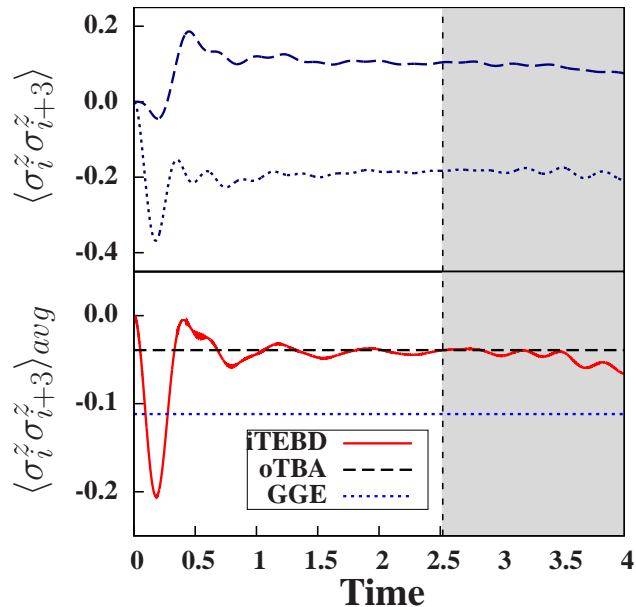}
\protect\caption{\label{zzThermalization} The upper panel shows the relaxation of
the $\langle\sigma_{i}^{z}\sigma_{i+3}^{z}\rangle$ correlator in
the two sublattices, calculated by the iTEBD algorithm. The simulation
was started from the dimer state and the anisotropy parameter was
set to $\Delta=4$. In the shaded region the simulation is unreliable
because of the truncation error. The translational invariance is not
restored in the achievable timescale of the iTEBD simulation. The
lower panel shows the relaxation of the sublattice averaged correlator
calculated by the iTEBD code (continuous red line). The blue dotted
line shows the GGE prediction, while the dashed black line represents
the oTBA prediction. The GGE result deviates strongly from the iTEBD
simulations. The oTBA prediction, however, describes the stationary
value with high precision.}
\end{figure}
The real time dynamics of the spin chains were numerically simulated
using the Infinite-size Time Evolving Block Decimation (iTEBD) algorithm
developed in \cite{2004PhRvL..93d0502V,2007PhRvL..98g0201V}. In this
algorithm the state of the system is approximated by a matrix product
state (MPS) \cite{2011AnPhy.326...96S}, and the time-evolution is
approximated in a Suzuki-Trotter expansion by consecutive two-site
evolution steps. In the calculations the rotational symmetry around
the $z$-axis was exploited by introducing the conserved charge --
the $z$-component of the total spin -- and performing the time steps
for the different charge blocks separately.

The precision of the MPS Ansatz can be controlled by the dimensions
of the matrices. The required sizes depend strongly on the entanglement
between separate parts of the system. In a real time simulation of
an infinite system, the entanglement of the simulated state grows
rapidly. The required MPS-dimension ($\chi$) grows exponentially
that results in a threshold, where the simulation loses its reliability.
Due to the exponential growth, the threshold time can only slowly
move by increasing $\chi$.

In our calculations maximal dimensions of the U(1) blocks were set
to $\chi_{block}=400-800$ resulting in a total matrix dimension $\chi_{tot}>1200-2400$.
In the time-evolution a first order Suzuki-Trotter expansion was used
with a time-step $dt=0.001$. We tested the reliability of discretization
by changing the time-step and found that decreasing $dt$ does not
modify our results.

The $\langle\sigma_{i}^{x}\sigma_{j}^{x}\rangle$ and $\langle\sigma_{i}^{z}\sigma_{j}^{z}\rangle$
correlations were calculated at each time step up to third neighbour
distance. Preparing the system in the dimer and q-dimer states --
in contrast with the Néel state -- these correlations are not invariant
under translations by the lattice constant. In the Fig. \ref{zzThermalization}
the $\langle\sigma_{i}^{z}\sigma_{i+3}^{z}\rangle$ correlator is
shown when the relaxation starts from the dimer state. The translational
invariance is not restored in the achievable simulation times and
it is still an open question whether it is restored in the $t\rightarrow\infty$
limit or not. The sublattice averaged correlator, however, converge
rapidly to a stationary value that is equal to the oTBA prediction
within the numerical errors.

\begin{figure}
\centering{}\includegraphics[width=0.48\textwidth]{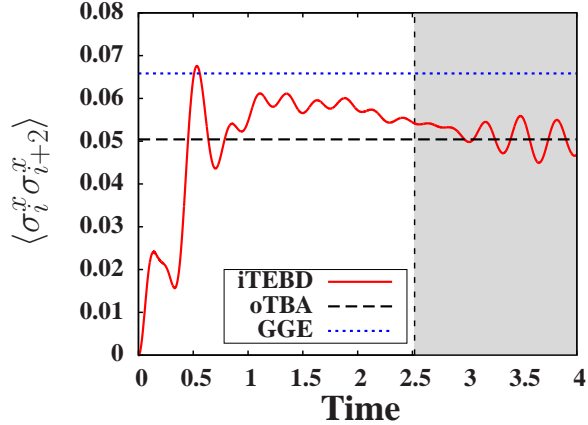}
\protect\caption{\label{xxThermalization} The relaxation of the $\langle\sigma_{i}^{x}\sigma_{i+2}^{x}\rangle$
correlator is shown (red continuous line). The simulation was started
from the dimer state, and the anisotropy parameter was set to $\Delta=4$.
The blue dotted line shows the GGE prediction, while the dashed black
line represents the oTBA prediction. Compared to the $zz$ correlator
(see Fig. \ref{zzThermalization}) the relaxation is much slower.
Because of the slow drift an accurate extrapolation for the stationary
value was not possible.}
\end{figure}

To extract the stationary correlators from the iTEBD simulations we
defined the threshold of the simulation at the time where the one-step
truncated weight exceeds $10^{-8}$. As an alternative method we ran
the code with different matrix dimensions and determined the time
where the results deviated upon increasing matrix dimension. The two
methods resulted approximately in the same threshold time. Because
we do not have any reliable analytic expression for the real-time
evolution of the correlators, the estimated stationary values were
calculated by simple time averaging for the last $\Delta T=1$ time
window before the threshold. The error bars were estimated by the
minimal and maximal values within the time window. The method works
well in the case of $zz$-correlators (see Fig. \ref{zzThermalization}),
where the correlator rapidly converge to the stationary value. In
that case the amplitude of small oscillations gives a reasonable error-bar.
In the case of $xx$-correlators, however, the convergence is much
slower (see Fig. \ref{xxThermalization}). The correlator slowly drifts
towards a stationary value but does not reach it before the threshold.
In the case of such a slow drift, the estimation of error bars with
minimal and maximal values in the time window is not optimal, because
it can not quantify how far the real stationary value is. 

\end{appendices}
\end{document}